\documentclass[onecolumn]{autart}

\newif\ifproofshp 
\proofshptrue

\usepackage{times} 
\usepackage{amsmath} 
\usepackage{amssymb}  
\usepackage{amsfonts}
\usepackage{cite}
\usepackage{graphics} 
\usepackage{graphicx}
\usepackage{breqn}
\usepackage{enumerate}
\usepackage{tikz}
\usetikzlibrary{decorations.markings}
\usepackage{ctable}

\newcommand{\PMP}{the Pontryagin Maximum Principle}
\newcommand{\pdr}{\dot{\phi}_R}
\newcommand{\pdl}{\dot{\phi}_L}
\newcommand{\pddr}{\ddot{\phi}_R}
\newcommand{\pddl}{\ddot{\phi}_L}
\newcommand{\torqr}{u_1}
\newcommand{\torql}{u_2}
\newcommand{\torqmax}{u_m}
\newcommand{\mysign}[1]{\mathrm{sign}(#1)}

\newcommand{\myPN}{\alpha^{-}}
\newcommand{\myNP}{\alpha^{+}}
\newcommand{\myNN}{\beta^{-}}
\newcommand{\myPP}{\beta^{+}}
\newcommand{\apmphase}{\alpha^{\pm}}
\newcommand{\ampphase}{\alpha^{\mp}}
\newcommand{\bpmphase}{\beta^{\pm}}
\newcommand{\bmpphase}{\beta^{\mp}}
\newcommand{\admissct}{\Pi}

\newcommand{\R}{\mathbb{R}}

\newcommand{\pd}[2]{\frac{\partial #1}{\partial #2}} 
 
\newcommand{\bmat}[1]{\begin{bmatrix}#1\end{bmatrix}}
\newcommand{\mc}[1]{\mathcal{#1}}
\newcommand{\bigpar}[1]{\left( #1 \right)}
\newcommand\adjointinout{}
\makeatletter

\newcommand{\Rmnum}[1]{\expandafter\@slowromancap\romannumeral #1@}
\makeatother

\begin{document}
\tikzset{->-/.style={decoration={  markings,  mark=at position .5 with{\arrow{>}}},postaction={decorate}}}
\tikzset{-<-/.style={decoration={  markings,  mark=at position .5 with{\arrow{<}}},postaction={decorate}}}
%
%


\begin{frontmatter}

\title{{Time-optimal Velocity Tracking Control for Differential Drive Robots}}


\author[utexas]{Hasan A Poonawala}\ead{hasanp@utexas.edu},    
\author[utdecs]{Mark W Spong}\ead{mwspong@utdallas.edu},               

\address[utexas]{Institute for Computational Engineering and Sciences\\University of Texas at Austin\\ 201 E 24 St\\Austin, TX 78712, USA}
\address[utdecs]{Erik Jonsson School of Engineering \& Computer Science\\University of Texas at Dallas\\800 W Campbell Rd\\Richardson, TX, 75252, USA}


\begin{abstract}
Nonholonomic wheeled mobile robots are often required to implement algorithms designed for holonomic kinematic systems. This creates a velocity tracking problem for the actual wheeled mobile robot. In this paper, we investigate the issue of tracking the desired velocity in the least amount of time, for a differential drive nonholonomic wheeled mobile robot. If the desired velocity is a constant, the Pontryagin Maximum Principle can be used to design a control. A control is designed for the cases when the wheel can be commanded speeds and torques. When the desired velocity is smoothly time-varying, we propose a hybrid structure and study its properties.
\end{abstract} 

\end{frontmatter}
\maketitle
\begin{abstract}
Nonholonomic wheeled mobile robots are often required to implement algorithms designed for holonomic kinematic systems. This creates a velocity tracking problem for the actual wheeled mobile robot. In this paper, we investigate the issue of tracking the desired velocity in the least amount of time, for a differential drive nonholonomic wheeled mobile robot. If the desired velocity is a constant, the Pontryagin Maximum Principle can be used to design a control. A control is designed for the cases when the wheel can be commanded speeds and torques. When the desired velocity is smoothly time-varying, we propose a hybrid structure and study its properties.
\end{abstract} 


\section{Introduction}
\label{sec:intro}
Differential drive systems are a popular choice for mobile robot platforms. This can be attributed largely to their ability to turn in place, which makes them ideal for navigation in cluttered environments. Another advantage is the simplicity of construction, especially when compared to holonomic wheeled mobile robots. The control of nonholonomic wheeled mobile robots has a long history~\cite{Brockett83,Ryan94,Bloch03,Kuipers11}, with the differential drive robot system being a common example. The most important controls problem typically considered for this robot is the point stabilization problem~\cite{Samson91} or the tracking of a reference trajectory~\cite{Jiang97,dAndrea92,Fierro95}. The point stablization problem is particularly interesting due to the impossibility of solving it using a smooth time-invariant feedback law~\cite{Brockett83}. 

In recent years the field of multi-robot coordination has been an active area of research. Control methods such as consensus algorithms and behaviour-based controls can achieve a wide variety of tasks. In general, these methods often consider single integrator dynamics, and the commanded control for each robot is a velocity in the plane. Such control laws can be implemented in an exact manner only on holonomic wheeled mobile robots. Further, consider a team of multiple differential drive robots that are to be operated by a human using some input device. Typically, the human may command a motion towards a particular direction. Depending on the headings of the robots, they may or may not be able to move in that direction instantaneously.

Thus, in this paper, we are concerned with controlling the planar velocity of the differential drive robot. The goal is to find controls that change the current velocity of the robot to some desired velocity in the plane as fast as possible. The effect of implementing such controls is to make the robots `appear' to be holonomic, with as small a  delay as possible in tracking of commanded velocities. Previous work on time-optimal control for the differential drive robot has focused on control of the robot's position~\cite{Reister94,Renaud97,VanLoock13}.

The contribution of this paper lies in applying \PMP\ to the differential drive robot with bounded torque inputs in order to derive time-optimal controls that drive the forward speed, heading angle and angular velocity to desired values. 

\section{Preliminaries}
In this section we describe the differential drive robot system and recount the Pontryagin Maximum Principle which will be appplied to this system.
\subsection{Differential Drive Robot}
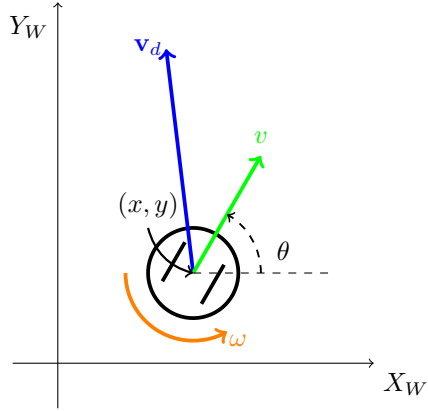
\begin{figure}[tb]
\centering
\begin{tikzpicture}[scale=0.6]
\draw[->] (-1,0) -- (7,0);
\draw[->] (0,-1) -- (0,8);
\draw (0,7) node[anchor=south east]{$Y_W$} ;
\draw (7,0) node[anchor=north west]{$X_W$} ;
\begin{scope}[shift={(3cm,2cm)},rotate=-30,]
\draw[line width = 0.5mm](0cm,0cm) circle (1cm);
\draw[line width = 0.5mm] (0.5cm,-0.5cm) -- (0.5cm,0.5cm);
\draw[line width = 0.5mm] (-0.5cm,-0.5cm) -- (-0.5cm,0.5cm);
\draw[->,color=blue,line width = 0.5mm] (0,0) -- (-3,4);
\draw[->,color=green,line width = 0.5mm] (0,0) -- (0,3);
\end{scope}
\draw[->,shift = {(1.5cm,2cm)},color=orange,line width = 0.5mm] (0cm,0cm) arc (180:300:1.5cm);
\draw (2cm,3cm) node[anchor = south]{$(x,y)$};
\draw[->,line width =0.3mm] (2cm,3cm) to [bend right] (3cm,2cm);
\draw[dashed] (3cm,2cm) -- (6cm,2cm);
\draw[->,shift = {(4.5cm,2cm)},dashed,line width = 0.25mm] (0cm,0cm) arc (0:60:1.5cm);
\draw (5,2.5) node{$\theta$};
\draw[color = blue] (2,7) node{$\mathbf{v}_d$};
\draw[color = green] (4.5,5) node{$v$};
\draw[color = orange] (4,0.5) node{$\omega$};
\end{tikzpicture}
\caption{The differential drive robot with with linear speed $v$, angular velocity $\omega$ and desired velocity $\mathbf{v}_d$.}
\label{fig:ddwmr}
\end{figure}

A sketch of a differential drive robot is shown in Figure \ref{fig:ddwmr}. The desired velocity $\mathbf{v}_d \in \R^2$ is given by the blue vector, with magnitude $v_d = \| \mathbf{v}_d\|$. However, the robot's velocity lies along the green vector, with magnitude $v \in \mathbb{R}$. The robot heading $\theta$ must be controlled such that the robot velocity matches the desired one. 

The kinematic equations of motion of the wheeled mobile robot are
\begin{dmath}
\bmat{\dot{x}\\\dot{y}\\ \dot{\theta}} = \bmat{\cos{(\theta)} & 0 \\ \sin{(\theta)} & 0 \\0 & 1} \bmat{v \\ \omega}
\label{eq:ddkinematics}
\end{dmath}
\noindent where $(x,y)$ is the cartesian position of the centroid of the robot, $v$ is the forward speed and $\omega$ is the angular velocity of the robot. The non-holonomic nature of the equations is due to the fact that the equations \ref{eq:ddkinematics} satisfy the contraint
\begin{dmath}
\dot{x} \sin{(\theta)} - \dot{y} \cos{(\theta)} = 0
\end{dmath}

We assume that the wheels do not slip. This corresponds to the two constraints
\begin{dgroup*}
\begin{dmath}
\dot{x} \cos{(\theta)} + \dot{y} \sin{(\theta)} + b \dot{\theta} = r \pdr 
\end{dmath}
\begin{dmath}
\dot{x} \cos{(\theta)} + \dot{y} \sin{(\theta)} - b \dot{\theta} = r \pdl 
\end{dmath}
\end{dgroup*}

The linear speed $v$ and angular velocity $\omega$ are then obtained from the right and left wheel velocities ($\dot{\phi}_R$ and $\dot{\phi}_L$ respectively) as
\begin{dgroup}
\begin{dmath}
v = \dot{\phi}_R \frac{r}{2} + \dot{\phi}_L \frac{r}{2}
\end{dmath}
\begin{dmath}
\omega = \frac{\dot{\phi}_R r}{2 b} - \frac{\dot{\phi}_L r}{2 b}
\end{dmath}
\label{eq:ddrkinematics}
\end{dgroup}
\noindent where $r$ and $2 b$ are the radii of the wheels and the distance between the wheels respectively. 

Some commercially available differential drive robots, such as the iRobot Create, can only be commanded wheel speeds. Further, the wheel speeds that can be obtained are bounded. That is, $| \dot{\phi}_R | \leq \dot{\phi}_{max}$ and  $| \dot{\phi}_L | \leq \dot{\phi}_{max}$ for some $\dot{\phi}_{max} > 0$. 

A second possibility is when the motors are torque controlled. Let $\torqr$ and $\torql$ be the net torques at the right and left wheels respectively. We assume that these torques are bounded, that is, $| \torqr | \leq \torqmax$ and $| \torql | \leq \torqmax$ for some $\torqmax > 0$. In this case, we can derive
\begin{dgroup}
\begin{dmath}
m \dot{v} = \frac{r}{2} \torqr +  \frac{r}{2} \torql
\end{dmath}
\begin{dmath}
J_r \dot{\omega}  = \frac{ r}{2 b} \torqr -  \frac{r}{2 b} \torql 
\end{dmath}
\label{eq:ddwmrdynamics}
\end{dgroup}
\noindent where $m$ is the effective mass of the robot and $J_r$ is the effective rotational inertia of the robot about the vertical axis through the center of the wheel base. The parameters $m$ and $J_r$ are functions of the robot and wheel parameters ( see \cite{Sarkar94} for details). Further, the right and left wheel speeds change according to the equations
\begin{dmath}
\bmat{c_1 & c_2 \\ c_2 & c_1} \bmat{\pdr \\ \pdl} = \bmat{\torqr \\ \torql}
\label{eq:wheelspeeddynamics}
\end{dmath}
\noindent where $c_1$ and $c_2$ are strictly positive constants which depend on the robot parameters ( See \cite{Sarkar94} for details). Note that $c_1$ and $ c_2 $ cannot be equal since equality requires the robot to have no rotational inertia about the vertical axis . Also $c_2 = 0 \iff m b^2 = I$ ( see \cite{Sarkar94} for details). 
 
We will use both models \eqref{eq:ddrkinematics} and \eqref{eq:ddwmrdynamics} separately to address the goals outlined in Section \ref{sec:problemstatement}.
\subsection{Time-optimal Control}
Consider a dynamical system consisting of state $q \in \R^n$. 
The dynamics are given by
\begin{dmath}
\dot{q} = f(q,u)
\label{eq:dynamicalsystem}
\end{dmath}
\noindent where $u \in U \subset \mathbb{R}^p$ is the control input and $f: \R^n \times U \rightarrow \R^n$ is a vector field on $\R^n$.
Consider an initial state $q_0 \in \R^n$ and a target state $q_d \in \R^n$. Assume that there exists some control $u(t)$ defined on $[0,t_f]$ such that the corresponding trajectory $q(t)$ defined on $[0,t_f]$ has the property that $q(0) = q_0$ and $q(t_f) = q_d$. The pair $(q(t),u(t))$, where $t$ is defined on $[0,t_f]$, is called a controlled trajectory. Out of all controlled trajectories that achieve the desired change of state, the time-optimal control problem consists of finding one for which the final time $t_f$ the least-possible. 

The Pontryagin Maximum Principle \cite{Schattler,Mauder,SussTang,Wu2000} can be used as a tool to find these time-optimal controlled trajectories, by specifying necessary conditions that they must satisfy. Any controlled trajectory meeting the necessary conditions is called an extremal. The time-optimal controls are a subset of the extremals, hence application of \PMP\ is a good first step to finding them. Further sufficiency conditions need to be applied in order to conclude that an extremal  is time-optimal. 

We can introduce the adjoint state $\psi \in \R^n$, and the Hamiltonian $H$ given by 
\begin{dmath}
H(q,\mu,\psi,u) = - \mu + \psi^T f(q,u)
\label{eq:hamiltonian}
\end{dmath}
\noindent where $\mu \in \{0,1\}$.

The principle states that:
\begin{thm}
Consider system \eqref{eq:dynamicalsystem}  with $U$ a compact subset of $\R^{p}$. Let there exist an adjoint state $\psi \in \R^{n}$, a Hamiltonian function $H$ given by \eqref{eq:hamiltonian}, an extremal denoted by the triple $(q^*(t),\psi^*(t),u^*(t))$ and the extremal Hamiltonian  $H^*(t) = H(q^*(t),\mu,\psi^*(t),u^*(t))$ defined on $t \in I = [0,t_f]$. Then the following are true
\begin{enumerate}
\item[N1] For all $t \in I $, $(\mu, \psi^*(t)) \neq 0$ holds.
\item[N2] For almost all $t \in I$, the adjoint state satisfies
\begin{dmath}
\dot{\psi} = - \pd{H}{q}(q^*(t),\mu,\psi^*(t),u^*(t))
\label{eq:costatedynamics}
\end{dmath}
\item[N3] For almost all $t \in I$, $u^*(t)$ satisfies
\begin{dmath}
H^*(t) = \max_{u \in U} H(q^*(t),\mu,\psi^*(t),u)
\end{dmath}
\item[N4] For almost all $t \in I$, $H^*(t) = 0$ holds.
\end{enumerate}
\end{thm}
\section{Problem Statement}
\label{sec:problemstatement}
The position of the centroid of the wheeled mobile robot at time $t$ is denoted by $(x(t),y(t)) \in \R^2$. We are given a desired velocity  $\mathbf{v}_d \in \R^2$. The goal is to design control strategies such that the derivative of the position matches the desired velocity, and that the convergence is achieved as fast as possible.
\begin{dmath}
\bmat{\dot{x}\\\dot{y}} \rightarrow \mathbf{v}_d
\end{dmath}
Given the robot kinematics \eqref{eq:ddrkinematics}, this is equivalent to requiring that 
\begin{dmath}
{\theta \rightarrow \theta_d,\ v \rightarrow \| \mathbf{v}_d \|}
\end{dmath}
\noindent where $\theta_d$ is the angle that $\mathbf{v}_d$ makes with the $x$-axis of the coordinate axis in which $(x,y)$ is defined.

We want to solve the control problem for two types of inputs:
\begin{enumerate}
\item The control inputs are the wheel (or motor) speeds
\item The control input are the wheel (or motor) torques
\end{enumerate}
\noindent and two types of reference signals
\begin{enumerate}
\item $\mathbf{v}_d$ is constant
\item $\mathbf{v}_d$ is time-varying
\end{enumerate}

In the rest of the paper, we address the case of wheel speed inputs for both types of reference signals, and torque inputs for the case of constant reference signal $\mathbf{v}_d$.


\section{Torque Control}
The state is taken as $q  = ( v, \theta, \omega)^T \in \R^3$, and its dynamics are given by \eqref{eq:ddwmrdynamics}. Note that $\theta$ is treated as a real number instead of an element of $\mathcal{S}^1$. The input space $U \subset \R^2$ is $[-\torqmax,\torqmax] \times [-\torqmax,\torqmax]$. We can write the dynamics in the form $\dot{q} = f(q,u)$ as shown:

\begin{dmath*}
\bmat{\dot{v}\\ \dot{\theta} \\ \dot{\omega}} = \bmat{\frac{r}{m} (\torqr + \torql) \\ \omega \\ \frac{2 r}{J_r b} ( \torqr - \torql) }
\end{dmath*}
Which is a linear system
\begin{align}
\bmat{\dot{v}\\ \dot{\theta} \\ \dot{\omega}} &= \bmat{0&0&0\\0&0&1\\0&0&0} q + \bmat{\frac{r}{m J}&\frac{r}{m J}\\0&0\\\frac{2 r}{J_r b}&-\frac{2 r}{J_r b}} \bmat{\torqr \\ \torql} \\
&= A q + B u
\label{eq:qdynamics}
\end{align}

We first check whether time-optimal controls that change the state from any initial state to any target state exist. We can use the Fillipov Existence Theorem~\cite{Hartl95,Mauder} to do this. 

\begin{thm}[Filippov Existence Theorem] Consider state $q \in \R^n$ with dynamics $\dot{q} = f(q, u)$, where $u \in U \subset \R^p$ and $U$ is compact. Time-optimal solutions exist if the control system is controllable, $f(q, u)$ satisfies the linear growth condition $\| f(q, u) \| \leq  c (1 + \|x\|)$ for some constant $c > 0$ and all $(q, u) \in \R^n \times U$, and the velocity sets $F_U(q) := \{f(q, u) | u \in¸ U\}$ are convex for all $q \in \R^n$.
\label{thm:fet}
\end{thm}

We can now show that our system does possess time-optimal controlled trajectories between any two states:
\begin{prop} There exists time optimal trajectories between any two state for the dynamical system \eqref{eq:qdynamics} with input space $U$.
\end{prop}
\begin{pf}
The system \eqref{eq:qdynamics} is a controllable linear system, which is trivial to check. The set of allowable inputs $U = [-\torqmax,\torqmax] \times [-\torqmax,\torqmax]$ is compact and convex. Thus, $B u$ is a convex subset of $\R^3$ and hence $A q + B u =  f(q,u)$ is convex for each $q \in \R^n$. The norm of $f(q,u)$ can be bounded as follows:
\begin{align*}
\| A q + B u \| & \leq \| A q \| + \| B u \|\\
& \leq \| q \| +  \|c\|\\
& \leq \max( \{1,\| c \| \}) ( 1 + \| q \|)
\end{align*}
\noindent where $c =  \bmat{2 \frac{r}{m}&0&\frac{4 r}{J_r b}}^T$ . Thus, $f(q,u)$ satisfies a linear growth condition. The system \eqref{eq:qdynamics} satisfies the conditions for Theorem \ref{thm:fet}, and hence time optimal controls that change any initial state to any target state exist.
\end{pf}

We can conclude that it makes sense to search for time-optimal controlled trajectories. We begin by constructing the extremals through application of \PMP. We assume that the extremals are defined over some compact time interval $I = [0,t_f]$. Given an extremal $(q^*(t),\psi^*(t),u^*(t))$, we refer to $q^*(t)$ as the  extremal trajectory and $u^*(t)$ as the extremal control. The adjoint state dynamics are derived using $N2$. The partial derivative $\pd{H}{q}$ is computed as: 
\begin{dmath}
\pd{H}{q} = \pd{\ }{q} \bigpar{ - \mu + \psi^T (A q + B u)}\\
 = A^T
\end{dmath}
The adjoint state dynamics are thus $\dot{\psi} = -A^T \psi$:
\begin{dmath}
\bmat{\dot{\psi}_1\\ \dot{\psi}_2 \\ \dot{\psi}_3} = \bmat{0 \\ 0 \\  -\psi_2}
\label{eq:quxnofriction}
\end{dmath}

The solution of this system given initial condition $\psi(0) = (\psi_1(0),\psi_2(0),\psi_3(0))$ is simply
\begin{dmath}
{\psi_1(t) = \psi_1(0), \psi_2(t) = \psi_2(0),\psi_3(t) = \psi_2(0) t + \psi_3(0) }
\label{eq:solnauxillary}
\end{dmath}

The Hamiltonian function becomes
\begin{dmath}
{H = - \mu +  \psi^T f(q,u) = - \mu + \psi^T A q + psi^T B u}\\
= - \mu +\psi_1 \bigpar{\frac{r}{ m} (\torqr + \torql)} + \psi_2 ( \omega ) + \psi_3 \bigpar{\frac{2 r}{J_r b} ( \torqr - \torql)} \\
 = \bigpar{\frac{r}{m} \psi_1 + \frac{2 r}{J_r b}  \psi_3 } \torqr + \bigpar{\frac{r}{m} \psi_1 - \frac{2 r}{J_r b}  \psi_3 }\torql + \psi_2 \omega
 \label{eq:hamiltonian2}
\end{dmath}
\subsection{Classification of extremals}
We are now in a position to determine extremals $(q^*(t),\psi^*(t),u^*(t))$. For any extremal, the function $H^*$ is maximized at each time instant $t$ (necessary condition $N3$), implying that
\begin{dgroup}
\begin{dmath}
\torqr(t) = \torqmax \ \mathrm{sign}\bigpar{\frac{r}{m} \psi_1(t) + \frac{2 r}{J_r b}  \psi_3(t) }
\end{dmath}
\begin{dmath}
\torql(t) = \torqmax \ \mathrm{sign}\bigpar{\frac{r}{m} \psi_1(t) - \frac{2 r}{J_r b}  \psi_3(t) }
\end{dmath}
\label{eq:pmptorques}
\end{dgroup}

The initial condition $\psi(0)$ determines $\psi^*(t)$ according to \eqref{eq:solnauxillary} , which in turn determines $u^*(t)$ according to \eqref{eq:pmptorques} and hence $q^*(t)$, given $q(0)$. The possible extremals are thus determined by the initial conditions $\psi(0)$. Clearly, three possibilities exist: 
\begin{enumerate}
\item $\psi(t) \equiv 0$
\item $\psi(t) \equiv \psi(0) \neq 0$
\item $\psi(t) \neq \psi(0)\ \forall t > 0$
\end{enumerate}

For convenience, we define the following \emph{switching functions}:
\begin{dgroup}
\begin{dmath}
\sigma_1(t)  = \bigpar{\frac{r}{m} \psi_1(t) + \frac{2 r}{J_r b}  \psi_3(t) }
\end{dmath}
\begin{dmath}
\sigma_2(t)  = \bigpar{\frac{r}{m} \psi_1(t) - \frac{2 r}{J_r b}  \psi_3(t) }
\end{dmath}
\label{eq:switchfuncs}
\end{dgroup}

\noindent \emph{Case 1:} If $\psi(t) \equiv 0$ then $H = -\mu + \psi_2 \omega$ and any control $u^*(t) \in U\ \forall t \in I$ would satisfy $N3$.  Such a case is known as the doubly-singular control. However, such a control cannot be an extremal control. This is due to the fact that $N1$ and $N4$ cannot simultaneously hold, and thus $\psi(t) \equiv 0$ cannot be part of a valid extremal. 

\noindent \emph{Case 2:} If $\psi(t) \equiv \psi(0) \neq 0$, which occurs when $\psi_2(0) = 0$. Consider either of the two mutually exclusive possibilities:
\begin{enumerate}
\item[$S1$] $\sigma_1(t) = \sigma_1(0) = 0$
\item[$S2$] $\sigma_2(t) = \sigma_2(0) = 0$
\end{enumerate}

When $S1$ ( $S2$ ) holds, the coefficient of $\torqr$ ($\torql$) in \eqref{eq:hamiltonian2} is zero, while the coefficient of $\torqr$ ($\torql$) is a non-zero constant. This implies that extremal controls may be of the form where one motor torque is $\pm \torqmax$ over the interval of definition of the trajectory, while the other control is arbitrary. 

Suppose $\psi(t) \equiv \psi(0) \neq 0$ however $S1$ and $S2$ do not hold. Then, according to \eqref{eq:pmptorques} the motor torques are constant with maximum possible magnitude. 

\noindent \emph{Case 3:} Finally, suppose that $\psi_2(0) \neq 0$, implying that $\psi(t) \neq \psi(0)\ \forall t > 0$. Since $\psi_1(t)$ is constant and $\psi_3(t)$ is linear in time $t$, $\sigma_1(t)$ ( or $\sigma_2(t)$) either monotonically increases of monotonically decreases, with exactly one time instant where its value is undefined. Since $\torqr = \torqmax \mysign{\sigma_1}$ and $\torql = \torqmax \mysign{\sigma_2}$, this implies that the motor torques are piecewise constant (with value $\pm \torqmax$) with no more than one switch.

To summarize, the application of \PMP\ results in the conclusion that all extremal controls consist of only two possible cases:
\begin{enumerate}
\item[$C1$] At least one motor has a constant torque with value $\torqmax$ or $-\torqmax$ over $I$.
\item[$C2$] Both motors have piecewise constant torques (with possible values in $\{-\torqmax,+\torqmax \}$) with exactly one switch for each motor at a time instants $t_1$ and $t_2$ such that $t_1$, $t_2 \in (0,t_f)$.
\end{enumerate}

We know that a time-optimal control between any two states exists. We also know that such a control must necessarily be of the form $C1$ or $C2$. Given a desired initial state $q_0$ and target state $q_d$, we attempt to find a control of the form $C1$ or $C2$ such that the control induces the desired change in state. This procedure generates an extremal $(q^*(t),\psi^*(t),u^*(t))$, defined on $t \in I = [0,t_f]$ such that $q(0) = q_0$ and $q(t_f) = q_d$. 

We now introduce a notation for the four possible combinations when both motor torques are at their maximum values. We name the combinations $\myPP$, $\myNN$, $\myNP$, and $\myPN$ as described in Table \ref{tab:motortorques}. When $\torqr = \torqmax$ and $\torql = -\torqmax$, for example, we refer to this situation as control being $\myNP$. For any interval of time where $\myPP$ or $\myNN$ control is used, the robot's linear speed changes while the angular velocity remains constant. For any interval of time where $\myNP$ or $\myPN$ control is used, the linear speed remains the same, and the angular velocity changes. Let the rates of angular acceleration and linear acceleration be $\alpha = 4 r \torqmax / ( J_r b)$ and $\beta = 2 r \torqmax / m$ respectively. Whenever the motors torques are equal to $\pm \torqmax$, either $\ddot{\theta} = \pm \alpha$ and $\dot{v} = 0$, or  $\ddot{\theta} = 0$ and $\dot{v} = \pm \beta$. 


\begin{table}
\caption{Notation for four torque modes}
\centering
\begin{tabular}{c|cc}
\hline
  &$ \torqr = \torqmax$ & $ \torqr = -\torqmax$ \\
 \hline
$\torql = \torqmax$ & $\myPP$ & $\myPN$\\
$\torql = -\torqmax$ & $\myNP$ & $\myNN$\\
\hline
\end{tabular}
\label{tab:motortorques}
\end{table}

For a $C2$ control, since each motor will switch exactly once, any $C2$ extremal consists of at most two instants of switching, and therefore at most three time intervals of time in which one of the four controls in Table \ref{tab:motortorques} is used. We call each time interval a phase of the extremal. We will refer to the controls used during any such interval using Table \ref{tab:motortorques}. If the control $u_1 = \torqmax$, $u_2 = -\torqmax$ is used during a phase, for example, we refer to that phase as a $+\alpha$ phase.
The possible sequences of control phases that are valid $C2$ extremal controls sequences are
\begin{enumerate}
\item $\apmphase$ $\rightarrow$ $\bpmphase$ $\rightarrow$  $\ampphase$
\item $\apmphase$ $\rightarrow$  $\ampphase$
\item $\bpmphase$ $\rightarrow$ $\apmphase$ $\rightarrow$  $\bmpphase$
\item $\bpmphase$ $\rightarrow$  $\bmpphase$
\end{enumerate}

\noindent where the arrow denotes a transition between one control phase (on the left of the arrow) to another control (on the right) at some time instant. Since both motors must switch exactly once, the control in the last phase is always the reversal of the control in the first phase.

Thus, we can introduce a further classification of the extremals, based on the possible combinations of motor torques listed above. The first two sequences are classified as $C2a$ controls. The last two are $C2b$ controls. The classification is based on whether the motor torques in the first phase have the same sign or not. 

For any $C1$ extremal control, one motor torque is always $+\torqmax$ or $-\torqmax$ and never switches during the transition from initial to goal state. The other motor torque can be an arbitrary function of time (bounded by $\torqmax$). Thus, extremal controls of the form $C1$ include singular controls, where one motor torque is arbitrary. We want to identify a special subset of $C1$ controls where the non-constant motor switches between $\pm \torqmax$ no more than twice. We will say that such controls are of type $C1_{ns}$. The possible $C1_{ns}$ extremal control sequences are
\begin{enumerate}
\item $\bpmphase$ $\rightarrow$ $\apmphase$ $\rightarrow$  $\bpmphase$
\item $\bpmphase$ $\rightarrow$ $\apmphase$
\item $\bpmphase$
\item $\apmphase$ $\rightarrow$ $\bpmphase$
\item $\apmphase$
\end{enumerate}

The next two subsections deal with the construction of extremal controls given $q_0$ and $q_d$, under the assumption that the desired angular velocity is zero. 

%
%
%

\subsection{Synthesis when $\omega(0) = 0$}
\label{ssec:synthwzero}
We are interested in target states where the robot has some desired velocity $\mathbf{v}_d$ in the plane. This corresponds to a desired forward speed $v_d = \| \mathbf{v}_d \|$ and orientation $\theta_d$, with zero angular velocity. In this subsection, we will focus on initial states where the robot angular velocity $\omega(0)$ is zero. Due to the fact that $f(q,u) = f(q + [v,\theta,0]^T,u)$, we can change coordinates such that the target state as $(0,0,0)$. The initial state $(v(0),\theta(0),0)$ becomes $(v_0,\theta_0,0)$ in the new coordinates. Thus, we are interested in transitioning from $(v0,\theta_0,0)$ to the origin $(0,0,0)$, the latter corresponding to $(v_d,\theta_d,0)$ in the original coordinates. Clearly, $v_0 = v(0) - v_d$ and $\theta_0 = \theta(0) - \theta_d$.

Assume that $\theta_0 = 0$. In order to avoid the trivial case, we must have $v_0 \neq 0$. Clearly, all we need to do is change the forward speed at the fastest possible rate to reach the origin. Thus, the extremal control is simply $\myPP$ or $\myNN$ for $t = \frac{|v|}{\beta}$ seconds. 
Note that the extremal control is of the form $C1a$.

Suppose that $\theta_0 \neq 0$. Since $\omega_0 = 0$, in order to change the robots heading from $\theta_0$, we must increase or decrease the angular velocity. However, since we wish to end with zero angular velocity, we must also decelerate by applying the opposite torques. This means that each motor switches exactly once, and hence we expect the extremal control that achieves the desired change in state to be of the form $C2a$.

Since there are two switches, we can divide the interval $[0,t_f]$ into three sub-intervals of length $t_1$, $t_2$ and $t_3$, where the motors switch at time instants $t_1$ and $\bar{t}_2 = t_1+t_2$ respectively. During the first and third interval, the control is of the form $\torqr = -\torql$. This does not clarify what the motor torques are, but merely that they are opposite in sign ($\myPN$ or $\myNP$). During the second the control is of the form $\torqr = \torql$ ($\myPP$ or $\myNN$).

The total duration of the trajectory is $\bar{t}_3 = t_1 + t_2 + t_3$. We can compute the final state at $\bar{t}_3$ due to a $C2a$ control through straightforward integration of the equations of motion as follows:
\begin{dgroup}
\begin{dmath}
\theta(\bar{t}_3) = \frac{s_1 \alpha t_1^2}{2}+ s_1 \alpha t_1 t_2  + s_1 \alpha t_1 t_3  -\frac{s_1 \alpha t_3^2}{2} + \theta_0
\label{eq:wzerothetat3}
\end{dmath}
\begin{dmath}
\omega(\bar{t}_3) = s_1 \alpha t_1 - s_1 \alpha t_3
\label{eq:wzerowt3}
\end{dmath}
\begin{dmath}
v(\bar{t}_3) = s_2 \beta t_2 +v_0
\label{eq:wzerovt3}
\end{dmath}
\label{eq:wzero}
\end{dgroup}
\noindent where $s_1 \in \{1,-1\}$ determines whether the first phase is $\myPN$ ($s_1 = -1$) or $\myNP$ ($s_1 = -1$), and $s_2   \in \{1,-1\} $ determines whether the second phase is $\myPP$ ($s_2 = 1$) or $\myNN$($s_2 = -1$). We can set $v(\bar{t}_3)=0$ in \eqref{eq:wzerovt3}, which results in the conclusion that
\begin{dgroup}
\begin{dmath}
t_2 = \frac{| v_0 |}{\beta}
\label{eq:wzerot2}
\end{dmath}
\begin{dmath}
s_2 = -\mysign{v_0}
\label{eq:wzeros2}
\end{dmath}
\label{eq:wzerot2s2}
\end{dgroup}

Similarily, setting  $\omega(\bar{t}_3)=0$ in \eqref{eq:wzerowt3} yields
\begin{dmath}
t_3 = t_1
\label{eq:wzerot3eqt1}
\end{dmath}

We can substitute \eqref{eq:wzerot2} and \eqref{eq:wzerot3eqt1} in \eqref{eq:wzerothetat3} along with the fact that we want $\theta(\bar{t}_3)=0$ to obtain
\begin{dmath}
s_1 \alpha t_1^2 + \frac{s_1 \alpha |v_0|}{\beta}  t_1 + \theta_0 = 0
\label{eq:wzerot1quadratic}
\end{dmath}
\noindent for which the solutions are 
\begin{equation}
t_{1,i} =  \frac{1}{2 \alpha} \bigpar{ -\frac{\alpha |v_0|}{\beta}   + (-1)^i  \sqrt{\frac{\alpha^2 |v_0|^2}{\beta^2} - \frac{4 \alpha \theta_0}{s_1} } }
\end{equation}
\noindent for $i \in \{1,2\}$. A non-negative solution always exists (when choosing $s_1 = -\mysign{\theta_0}$) which is given by
\begin{dmath}
t_1 =    \frac{1}{2 \alpha} \sqrt{\frac{\alpha^2 |v_0|^2}{\beta^2} + 4 \alpha |\theta_0 | } -\frac{|v_0|}{2 \beta} 
\end{dmath}
We can the compute the total time $\bar{t}_3 = t_1 + t_2 +  t_3$ as
\begin{dmath}
\bar{t}_3 =    \frac{1}{2 \alpha} \sqrt{\frac{\alpha^2 |v_0|^2}{\beta^2} + 4 \alpha |\theta_0 | } 
\end{dmath}

The switching times for the motors are $t_1$ and $t_1 + t_2$ respectively. The phases of the motor torques are determined by the $v_0$ and $\theta_0$, as described in Table \ref{tab:motorsequences}. 
 
\begin{table}
\centering
\caption{Control phases of extremals when $\omega(0) = 0$. A blank entry implies that that phase is non-existent.}
\begin{tabular}{ccccc}
\hline
$\theta_0$ & $v_0$ & $ 0 < t <t_1 $ & $t_1 < t <\bar{t}_2 $ & $ \bar{t}_2 <t < \bar{t}_3$  \\
\hline
$<0$ & $ < 0 $ & $\myNP$  & $\myPP$ & $\myPN $ \\
$<0$ & $ > 0 $ & $\myNP$  & $\myNN$ & $\myPN $ \\
$<0$ & $ = 0 $ & $\myNP$  & - & $\myPN $ \\
$>0$ & $ < 0 $ & $\myPN$  & $\myPP$ & $\myNP $ \\
$>0$ & $ > 0 $ & $\myPN$  & $\myNN$ & $\myNP $ \\
$>0$ & $ = 0 $ & $\myPN$  & - & $\myNP $ \\
$=0$ & $ < 0 $ & -  & $\myPP$ & - \\
$=0$ & $ > 0 $ & - & $\myNN$  & - \\
\hline
\end{tabular}
\label{tab:motorsequences}
\end{table}

\begin{figure}
\centering
\includegraphics[width=0.4\textwidth]{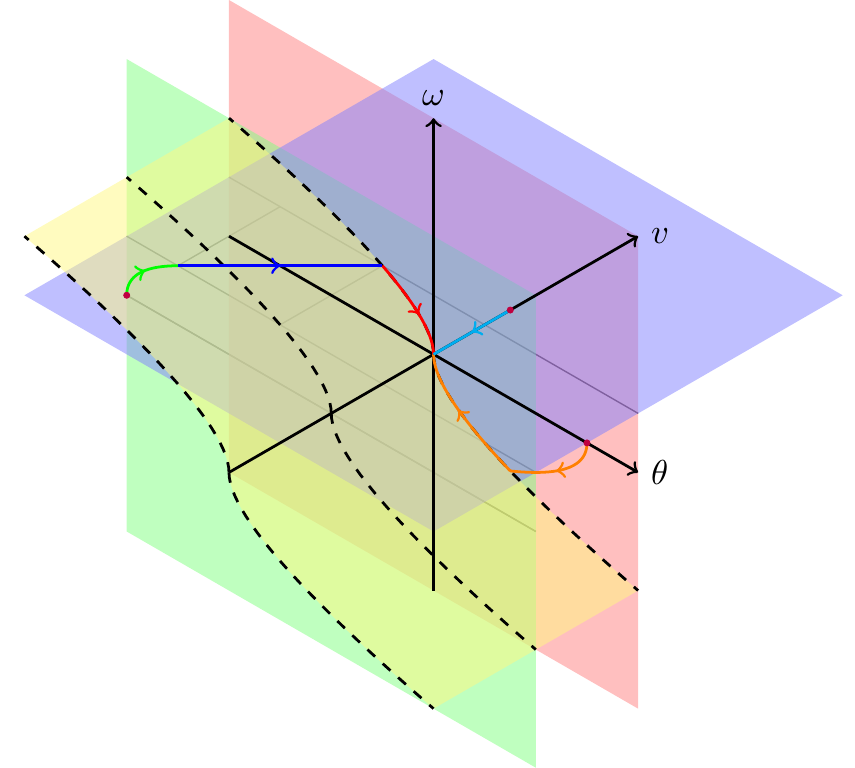}
\caption{Three optimal trajectories starting from the purple dots, marked by colored lines. The first trajectory consists of the green, blue and red curves. The second trajectory corresponds to the orange curve, for which there is no desired change in linear speed. The third trajectory corresponds to the case when $\theta_0=0$, and is represented by a cyan line. For the initial condition where $\theta_0 < 0$, the first phase of the trajectory represented by the green line lies in the (vertical) green plane $v = v_0$. The second phase of the trajectory represented by the blue line lies in a (horizontal) blue plane $\omega = c$, where $c$ is some constant. The third phase of the trajectory represented by the red line lies in the (vertical) red plane $v = 0$, as does the orange trajectory.}
\label{fig:maybe3d2}
\end{figure}

We can plot a sketch of these extremals in the $\R^3$, as done in Figure \ref{fig:maybe3d2}. Three extremals starting from the initial conditions marked by the three purple dots are seen in this figure. All three initial conditions are such that $\omega(0) = 0$. For the initial condition where $\theta_0 < 0$ the target linear speed is different from the initial linear speed. The trajectory consists of a sequence of phases $\myNP$, $\myPP$ (or $\myNN$ ), and $\myPN$. These are indicated by the green, blue and red lines respectively. For the initial condition where $\theta_0 > 0$ the target linear speed is identical to initial linear speed. As such, there are only two control phases: $\myPN$ and $\myNP$ (equivalently, $t_2 = \frac{| v_0 |}{\beta} =  0$). The case where only the linear speed needs to be changed is indicated by the cyan line. 

\begin{figure}
\centering
\begin{tikzpicture}[scale = 1]
\draw[->] (-4,0) -- (4,0);
\draw[->] (0,-2.5) -- (0,2.5);
\draw (4cm,0cm) node[anchor = north east]{$\theta$};
\draw (0cm,2.5cm) node[anchor = north east]{$v$};
\draw[->-,blue,thick] (-3,-2) -- (-1,0) ;
\draw[->-,green,thick] (-4,-2) -- (-3,-2) ;
\draw[->-,red,thick] (-1,0) -- (0,0) ;
\draw[->-,orange,thick] (3,0) -- (1.5,0) ;
\draw[->-,orange,thick] (1.5,0)-- (0,0) ;
\draw[->-,cyan,thick] (0,1.5)-- (0,0) ;
\draw[color = purple,line width = 0.1mm,fill = purple] (-4,-2) circle (0.05cm);
\draw[color = purple,line width = 0.1mm,fill = purple] (3cm,0) circle (0.05cm);
\draw[color = purple,line width = 0.1mm,fill = purple] (0,1.5) circle (0.05cm);
\end{tikzpicture}
\caption{Projection of the trajectories in Figure \ref{fig:maybe3d2} on to the plane $\omega=0$.}
\label{fig:vthetaplane}
\end{figure}
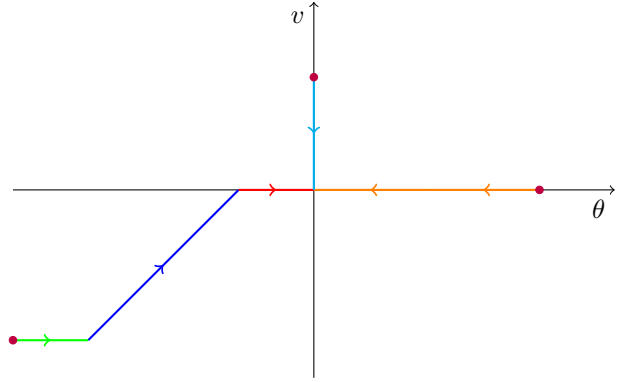

These same trajectories are projected on to the $\theta-\omega$ and $\theta -v$ planes in Figures \ref{fig:omegathetaplane} and \ref{fig:vthetaplane} respectively for more clarity. In Figure \ref{fig:omegathetaplane}, the dashed line corresponds to points where a single control phase $\myPN$ or $\myNP$ would be sufficient to reach the origin. The case when $\theta_0 = 0$ is plotted as a cyan dot at the origin in the $\theta - \omega$ plane. Note for that case when $v_0 = 0$, the problem reduces to time-optimal trajectories for the double integrator, with the target state being the origin \cite{Schattler}. 

\begin{figure}
\centering
\begin{tikzpicture}[scale = 1]
\draw[->] (-4,0) -- (4,0);
\draw[->] (0,-2.5) -- (0,2.5);
\draw (4cm,0cm) node[anchor = north east]{$\theta$};
\draw (0cm,2.5cm) node[anchor = north east]{$\omega$};
\begin{scope}[rotate=90]
\draw[dashed,thick] (0,0) parabola (2,4);
\draw[dashed,thick] (0,0) parabola (-2,-4);
\draw[-<-,blue,thick] (1,1) -- (1,3) ;
\end{scope}
\draw[->-,shift={(-4cm,0cm)},rotate=90,green,thick, domain=0:1] plot (\x, {-\x*\x});
\draw[-<-,rotate=90,red,thick, domain=0:1] plot (\x, {\x*\x});
\draw[-<-,shift={(0cm,0cm)},rotate=-90,orange,thick, domain=0:1.2] plot (\x, {\x*\x});
\draw[->-,shift={(2.88cm,0cm)},rotate=-90,orange,thick, domain=0:1.2] plot (\x, {-\x*\x});
\draw[color = purple,line width = 0.1mm,fill = purple] (-4,0) circle (0.05cm);
\draw[color = purple,line width = 0.1mm,fill = purple] (2.88cm,0) circle (0.05cm);
\draw[color = cyan,line width = 0.1mm,fill = cyan] (0,0) circle (0.05cm);
\end{tikzpicture}
\caption{Projection of the trajectories in Figure \ref{fig:maybe3d2} on to the plane $v=0$. The trajectory which corresponds to the case when $\theta_0=0$ gets projected to a point at the origin, represented by the cyan dot. The dashed curve represents points $(0,\theta,\omega)$ which would reach the origin if only the control $\myPN$ or $\myNP$ was used, for a suitable finite time period.}
\label{fig:omegathetaplane}
\end{figure}
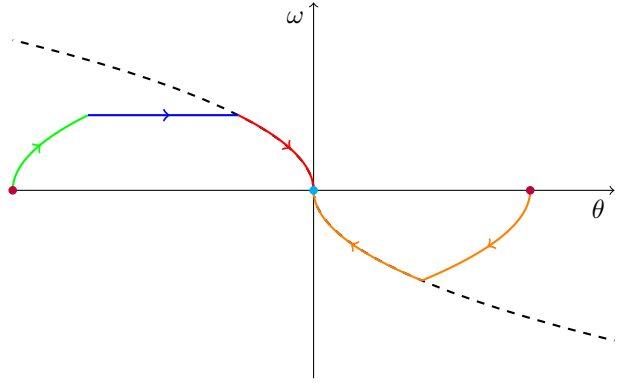

In summary, for the case when the robot must change from moving with one constant velocity in the plane to some other desired velocity in the plane, the extremal control is of the form $C2a$ unless the desired heading is the same as the initial heading, for which case the extremal control is of the form $C1_{ns}$. The switching times for the $C2a$ control have been derived.

\subsection{Synthesis for $\omega(0) \neq 0$}

In the previous subsection, we have found controls that satisfy the necessary conditions of \PMP\ when the initial angular velocity is zero, and these controls result in $v \rightarrow v_d, \theta \rightarrow \theta_d$. These trajectories can be found for any $\theta_0$ and $v_0$. More precisely, we can specify the motor torques as functions of time that achieve the change in state in the least possible time.

The case when $\omega(0) = 0$ corresponds to the mobile robot moving with a constant heading in the plane. We would like to accommodate the case when $\omega(0) \neq 0$ for various reasons, listed below:
\begin{itemize}
\item The robot could have been following a circular trajectory when a new desired linear velocity is commanded
\item Due to disturbances on the input or state, the robot may need to compute new switching times for an initial condition that corresponds to $\omega(0) \neq 0$ 
\item We wish to develop a state-based feedback control law that is time-optimal.
\end{itemize}

In Section \ref{ssec:synthwzero} we have found the time optimal control for initial conditions of the form $(v_0,\theta_0,0)$. For each such initial condition, there is a unique extremal $(q^*(t),\psi^*(t),u^*(t))$ defined on $I = [0,t_f]$ such that the initial condition is $q^*(0)$ and $q^*(t_f) = (0,0,0)$. Thus, this extremal trajectory is clearly the time-optimal one. 

In this subsection, we will see that we can derive more than one extremal for some initial conditions, and hence we need to establish which one is time-optimal when possible. Our strategy will be to determine sets of initial conditions where extremal controls of type $C1_{ns}$, $C2a$ or $C2b$ are such that the corresponding extremal trajectories start at the initial condition and end at the origin.

Let the initial condition be $q_0 = (v,\theta,\omega)$. We can define the following subsets of $\R^3$:

\begin{dmath}
\Omega_1 = \left\{ (v,\theta,\omega) \hiderel{\in} \R^3 \hiderel{:}   H_1(v,\theta,\omega) \hiderel{<} 0 \textrm{ and } {H_2(v,\theta,\omega) \hiderel{<} 0 }\right\}
\end{dmath}

\begin{dmath}
\Omega_2 = \left\{ (v,\theta,\omega) \hiderel{\in} \R^3 \hiderel{:}   H_1(v,\theta,\omega) \hiderel{>} 0 \textrm{ and } {H_2(v,\theta,\omega) \hiderel{>} 0 }\right\}
\end{dmath}

\begin{dmath}
\Omega_3 = \left\{ (v,\theta,\omega) \hiderel{\in} \R^3 \hiderel{:} \omega H_1(\theta,\omega) \hiderel{<} 0 \right\}
\end{dmath} 

\begin{equation}
\Omega_4 = \left\{ (v,\theta,\omega) \hiderel{\in} \R^3 \hiderel{:} H_1(v,\theta,\omega) H_2(v,\theta,\omega) \hiderel{<} 0 \right\}
\end{equation} 

\begin{dmath}
S_5 = \left\{ (v,\theta,\omega) \hiderel{\in} \R^3 \hiderel{:} H_1(v,\theta,\omega) \hiderel{=} 0, H_2(v,\theta,\omega) \hiderel{\neq} 0 \right\}
\end{dmath} 

\begin{dmath}
S_{6} = \left\{ (v,\theta,\omega) \hiderel{\in} \R^3 \hiderel{:} H_1(v,\theta,\omega) \hiderel{\neq} 0, H_2(v,\theta,\omega) \hiderel{=} 0 \right\}
\end{dmath} 

\begin{dmath}
L_v = \left\{ (v,\theta,\omega) \hiderel{\in} \R^3 \hiderel{:} H_1(v,\theta,\omega) \hiderel{=}  H_2(v,\theta,\omega) \hiderel{=} 0, v \hiderel{\neq} 0 \right\}
\end{dmath} 

\begin{dmath}
L_\omega = \left\{ (v,\theta,\omega) \hiderel{\in} \R^3 \hiderel{:} H_1(v,\theta,\omega) \hiderel{=}  H_2(v,\theta,\omega) \hiderel{=} 0, \omega \hiderel{\neq} 0 \right\}
\end{dmath} 

\noindent where 
\begin{dmath}
H_1(v,\theta,\omega) = 2 \alpha \theta + \omega |\omega| 
\end{dmath}
\noindent and
\begin{dmath}
H_2(v,\theta,\omega) = \frac{\omega | \omega | }{2 \alpha } + \theta + \frac{\omega |v| }{\beta}
\end{dmath}

\begin{figure}
\centering
\includegraphics[width=0.4\textwidth]{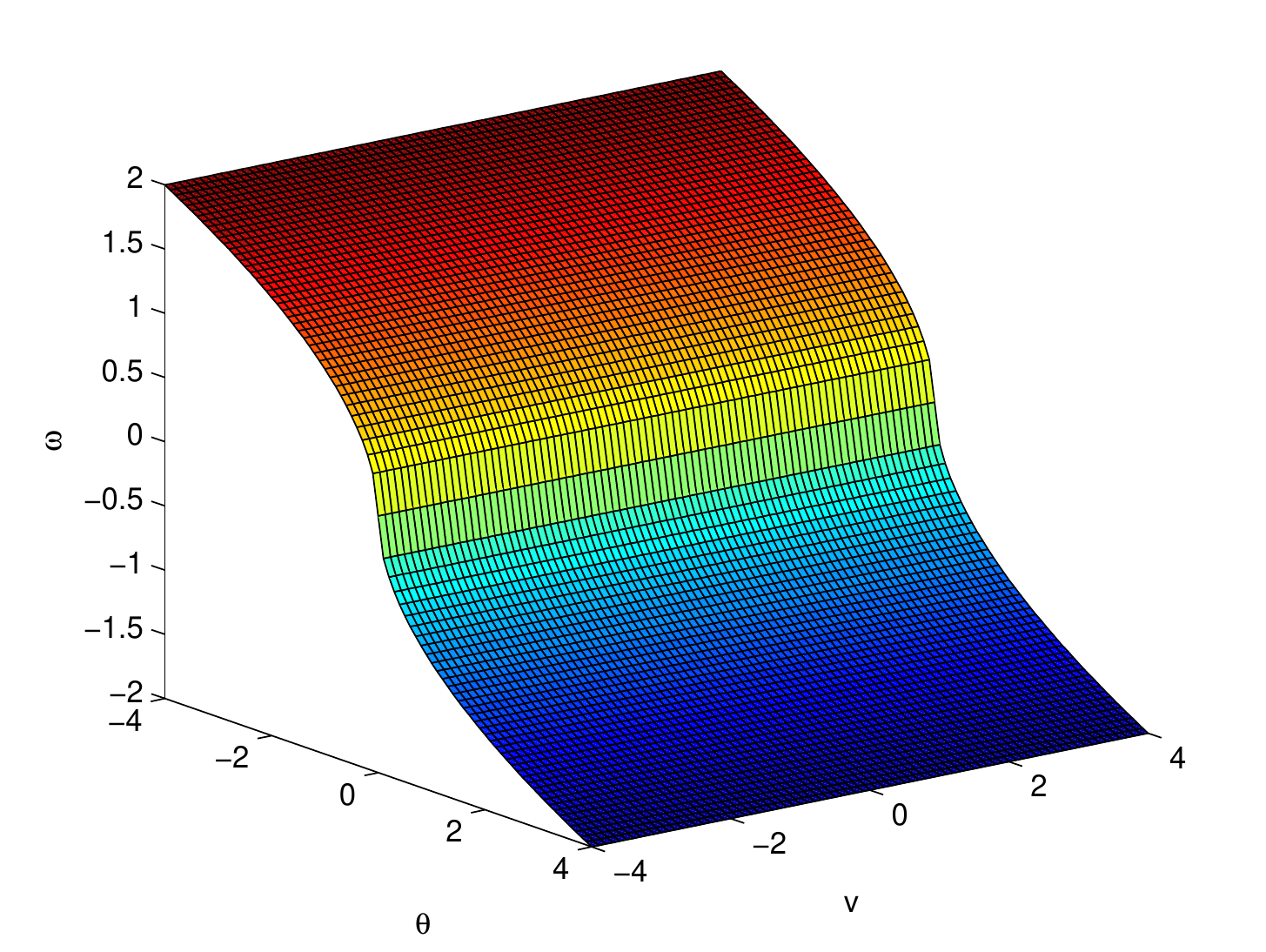}
\caption{The surface  $H_1(v,\theta,\omega) = 0$ for $\alpha = 0.5, \beta = 1$.}
\label{fig:H1eq0}
\end{figure}

\begin{figure}[t]
\centering
\includegraphics[width=0.4\textwidth]{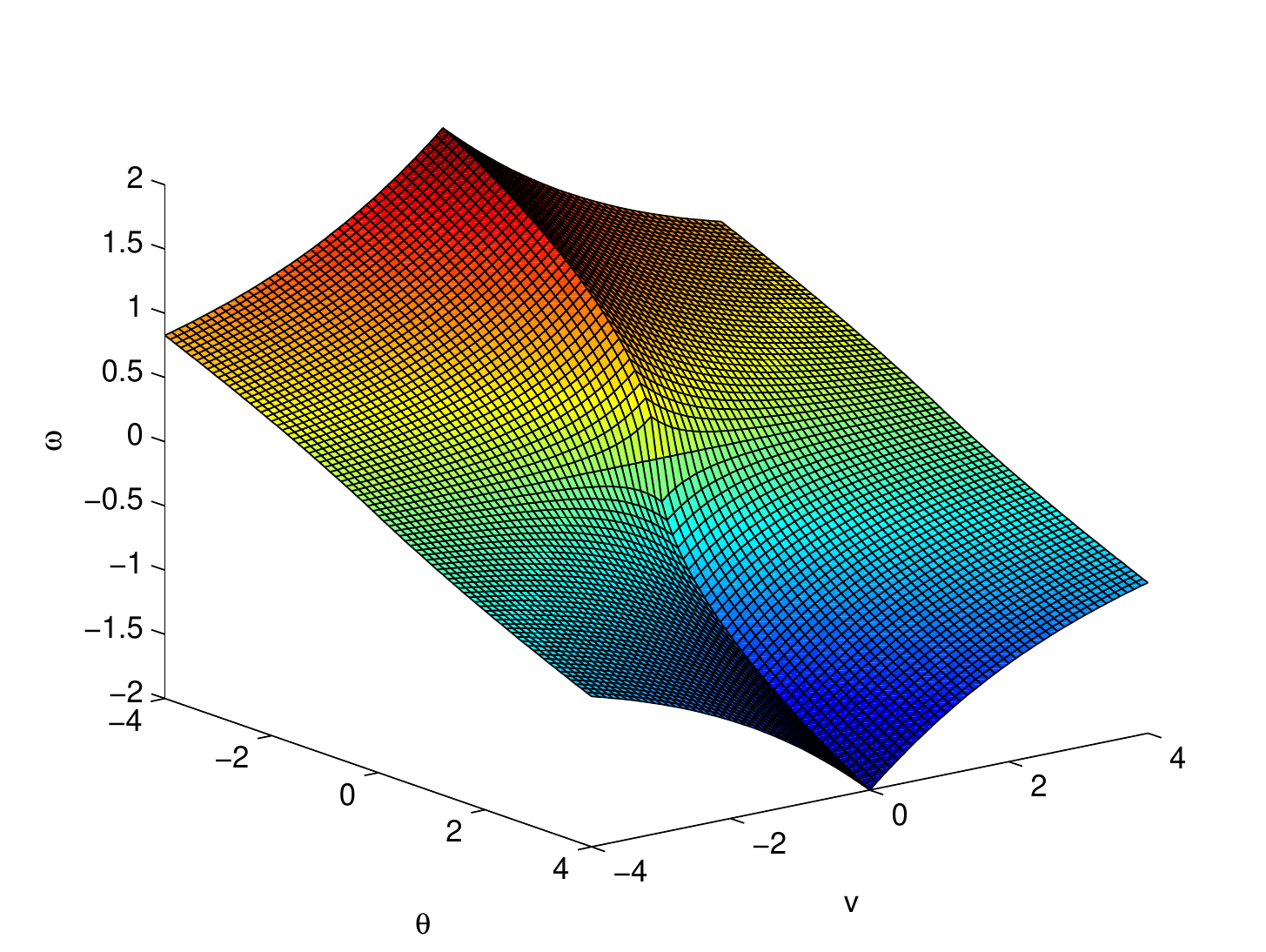}
\caption{The surface  $H_2(v,\theta,\omega) = 0$ for $\alpha = 0.5, \beta = 1$.}
\label{fig:H2eq0}
\end{figure}

\begin{figure}[t]
\centering
\includegraphics[width=0.4\textwidth]{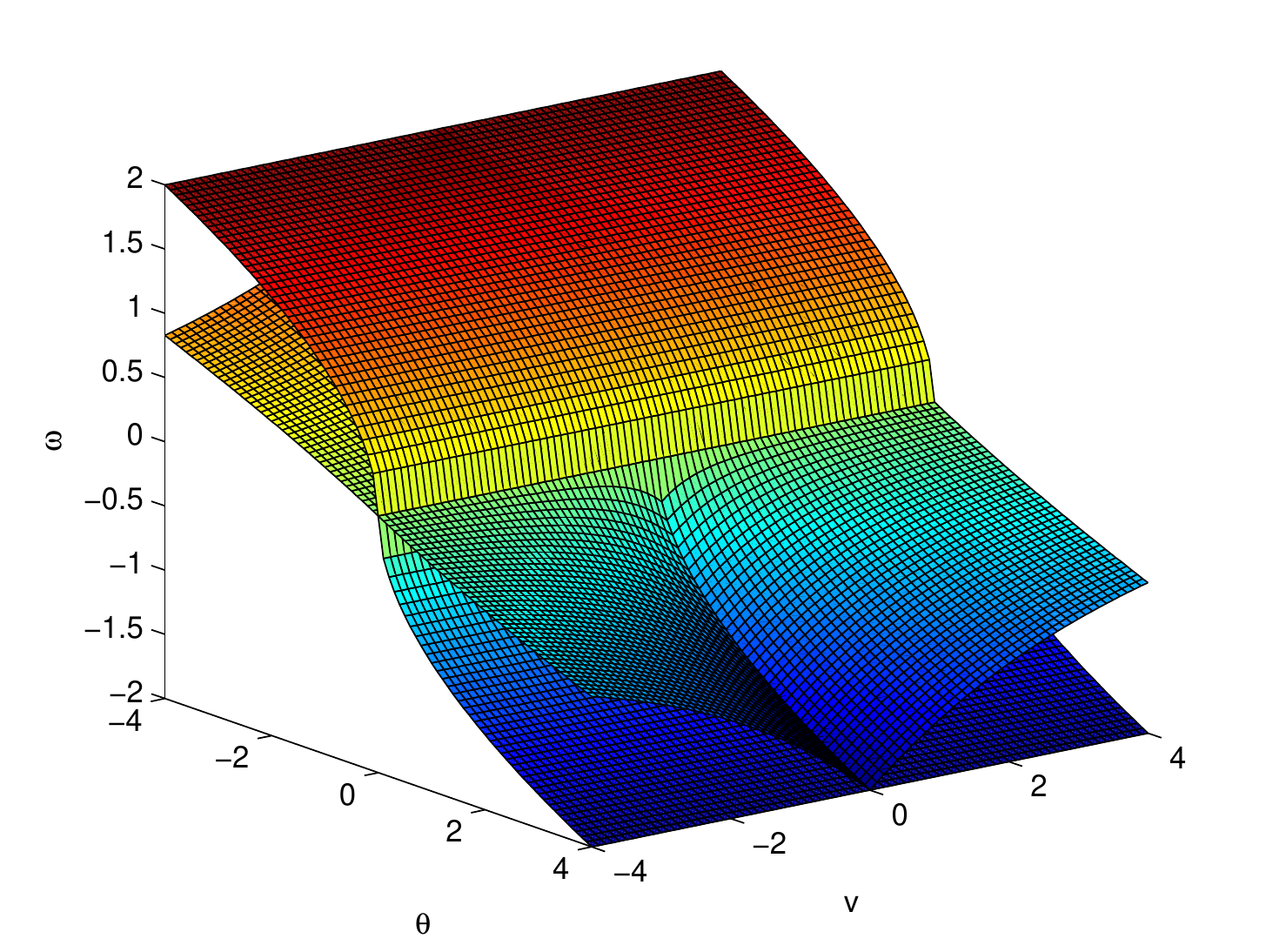}
\caption{The surfaces  $H_1(v,\theta,\omega) = 0$ and  $H_2(v,\theta,\omega) = 0$ for $\alpha = 0.5, \beta = 1$.}
\label{fig:H1H2eq0}
\end{figure}

The surfaces $H_1(v,\theta,\omega) = 0$ and $H_2(v,\theta,\omega) = 0$ are plotted in Figures \ref{fig:H1eq0} and \ref{fig:H2eq0} respectively. For comparison, they are superimposed in Figure \ref{fig:H1H2eq0}

\begin{lem}
Let $q_0 \in \Omega_1 \cup \Omega_2$. Then, an extremal $(q^*(t),\psi^*(t),u^*(t))$ defined on $I = [0,t_f]$ exists such that $q^*(0) = q_0$ and $q^*(t_f) = (0,0,0)$, and $u^*(t)$ is of type $C2a$.
\label{lem:omega12}
\end{lem}

\begin{pf}
Again, integration of the equations of motion for a general control of the form $C2a$ yields 
\begin{dgroup}
\begin{dmath}
\theta(\bar{t}_3) = \theta + \omega t_1 + \frac{s_1 \alpha t_1^2}{2}+ (\omega + s_1 \alpha t_1) t_2  + (\omega + s_1 \alpha t_1) t_3  -\frac{s_1 \alpha t_3^2}{2}
\label{eq:wnonzerothetat3}
\end{dmath}
\begin{dmath}
\omega(\bar{t}_3) = s_1 \alpha t_1 - s_1 \alpha t_3 + \omega
\label{eq:wnonzerowt3}
\end{dmath}
\begin{dmath}
v(\bar{t}_3) = s_2 \beta t_2 +v
\label{eq:wnonzerovt3}
\end{dmath}
\label{eq:wnonzero}
\end{dgroup}
\noindent which reduces to \eqref{eq:wzero} when $\omega = 0$. Again, $t_1$ and $\bar{t}_2 = t_1+t_2$ are the switching times of the motors, and $\bar{t}_2 = t_1+t_2+t_3$ is the total time. Also, $s_1$ and $s_2$ denote the unknown signs of the motor torques in the three phases.

The solutions for $t_2$ and $s_2$ remain unchanged from those in \eqref{eq:wzerot2s2}. Using \eqref{eq:wnonzerowt3} and requiring that $\omega(\bar{t}_3) = 0$ we can obtain
\begin{dmath}
t_3 = t_1 + \frac{\omega}{s_1 \alpha}
\label{eq:wnonzerowt3t1}
\end{dmath}
\noindent and hence we obtain the following quadratic equation in $t_1$:
\begin{dmath}
s_1 \alpha t_1^2 + \bigpar{2 \omega +\frac{s_1 \alpha |v|}{\beta}}  t_1 + \theta + \frac{\omega |v| }{\beta} + \frac{\omega^2}{2 s_1 \alpha}= 0
\label{eq:wnonzerot1quadratic}
\end{dmath}
\noindent for which the solutions are 
\begin{dmath}
t_{1,i} =  \frac{1}{2 s_1 \alpha} \bigpar{ -  2 \omega -  \frac{s_1 \alpha |v|}{\beta}   + (-1)^i  \sqrt{2 \omega^2 +\frac{\alpha^2 |v|^2}{\beta^2} - 4 s_1 \alpha \theta } }
\label{eq:wnonzerot1solncand}
\end{dmath}

From \eqref{eq:wnonzerowt3t1} we can obtain
\begin{dmath}
t_{3,i} =  \frac{1}{2 s_1 \alpha} \bigpar{ -  \frac{s_1 \alpha |v|}{\beta}   + (-1)^i  \sqrt{2 \omega^2 +\frac{\alpha^2 |v|^2}{\beta^2} - 4 s_1 \alpha \theta } }
\label{eq:wnonzerowt3solncand}
\end{dmath}

If a $C2a$ extremal control is to exist such that it drives the state from $q_0$ to the origin, we must have $t_1>0$ and $t_3>0$. Further, we can always take the solution corresponding to $i=2$, since $t_{3,1}$ can never be real and positive.

\noindent \emph{Case 1:} Consider the case when $s_1 = 1$. We have 
\begin{dgroup}
\begin{dmath}
t_{1} =  \frac{1}{2 \alpha} \bigpar{ -  2 \omega -  \frac{ \alpha |v|}{\beta}   +  \sqrt{2 \omega^2 +\frac{\alpha^2 |v|^2}{\beta^2} - 4  \alpha \theta } }
\end{dmath}
\begin{dmath}
t_{3} =  \frac{1}{2 \alpha} \bigpar{ -  \frac {\alpha |v|}{\beta}   +  \sqrt{2 \omega^2 +\frac{\alpha^2 |v|^2}{\beta^2} - 4 \alpha \theta } } 
\end{dmath}
\end{dgroup}

In order to obtain real solutions, we need the discriminant to be non-negative. This is achieved when
\begin{dmath*}
\omega^2 - 2 \alpha \theta \geq 0
\end{dmath*}
In order for to ensure that $t_3 > 0$, we further need
\begin{dmath}
\omega^2 - 2 \alpha \theta > 0
\label{eq:omega1condition1}
\end{dmath}

When $-  2 \omega -  \frac{ \alpha |v|}{\beta} >0$ this also implies that $t_1 >0$. When $-  2 \omega -  \frac{ \alpha |v|}{\beta} <0$, we need a further condition to hold in order that $t_1 >0$, given by

\begin{dmath}
 \alpha \theta + \frac{\alpha \omega |v| }{\beta} + \frac{\omega^2}{2} < 0
 \label{eq:omega1condition2}
\end{dmath}
\noindent which is obtained from the fact that for a quadratic equation $a x^2 + b x + c = 0$ where $b >0$, then one of the roots is real and positive if and only if $a c < 0$. 
We now show that the above conditions on $(v,\theta,\omega)$ hold when $H_1 < 0$ and $H_2 <0$. We will consider three cases, which depend on the value of $\omega$. 

\noindent \emph{Case 1a:} Let $\omega >0$. Then, 
\begin{align*}
& H_2(v,\theta,\omega) < 0 \\
\implies& \alpha H_2(v,\theta,\omega) < 0\\
\implies& \alpha \theta + \frac{\alpha \omega |v| }{\beta} + \frac{\omega | \omega |}{2} < 0\\
\implies& \alpha \theta + \frac{\alpha \omega |v| }{\beta} + \frac{\omega^2}{2} < 0
\end{align*}
\noindent and $\omega > - \frac{ \alpha |v|}{2 \beta}$.
Additionally,
\begin{align*}
&\alpha \theta + \frac{\alpha \omega |v| }{\beta} + \frac{\omega | \omega |}{2} < 0\\
\implies& \alpha \theta < - \frac{\alpha \omega |v| }{\beta} - \frac{\omega | \omega |}{2}\\
\implies& \alpha \theta < 0 < \frac{\omega^2}{2}\\
\implies& \omega^2 - 2 \alpha \theta > 0
\end{align*}

\noindent \emph{Case 1b:} Let $ - \frac{ \alpha |v|}{2 \beta} < \omega < 0$. We have 
\begin{align*}
&H_1 < 0\\
\implies& \omega | \omega | + 2 \alpha \theta < 0\\
\implies& - \omega^2 + 2 \alpha \theta < 0\\
\implies& \omega^2 - 2 \alpha \theta > 0
\end{align*}
Now, $ - 2 \omega - \frac{ \alpha |v|}{ \beta} <0$ which implies that $ 2 \omega^2 + \frac{ \alpha \omega |v|}{ \beta} < 0$ since $\omega >0$. We can add this to the negative quantity $-\frac{\omega^2}{2} + \alpha \theta $ to obtain
\begin{align*}
& 2 \omega^2 + \frac{ \alpha \omega |v|}{ \beta}  -\frac{\omega^2}{2} + \alpha \theta < 0\\
\Rightarrow&  \frac{\omega^2}{2} + \alpha \theta + \frac{ \alpha \omega |v|}{ \beta}  < - \omega^2 < 0
\end{align*}

\noindent \emph{Case 1c:} Let $ \omega <- \frac{ \alpha |v|}{2 \beta} < 0$. Once again $H_1 <0$ immediately implies that $\omega^2 - 2 \alpha \theta > 0$. Thus, $H_1 <0$ and $H_2 < 0$ implies that a $C2a$ control exists with first phase $PN$ such that the controlled trajectory reaches the origin. 

\noindent \emph{Case 2:} Let $s_1 = -1$, corresponding to the first phase being $NP$. The only valid solution to \eqref{eq:wnonzerot1quadratic} is given by
\begin{dgroup}
\begin{dmath}
t_{1} =  \frac{1}{2 \alpha} \bigpar{   2 \omega -  \frac{ \alpha |v|}{\beta}   +  \sqrt{2 \omega^2 +\frac{\alpha^2 |v|^2}{\beta^2} + 4  \alpha \theta } }
\end{dmath}
\begin{dmath}
t_{3} =  \frac{1}{2 \alpha} \bigpar{ -  \frac {\alpha |v|}{\beta}   + \sqrt{2 \omega^2 +\frac{\alpha^2 |v|^2}{\beta^2} + 4 \alpha \theta } } 
\end{dmath}
\end{dgroup}

Clearly, for $t_3 >0$ we need
\begin{dmath}
\omega^2 + 2 \alpha \theta > 0
\label{eq:omega2condition1}
\end{dmath} 
Again, this condition is sufficient to ensure $t_1 >0$ when $  2 \omega -  \frac{ \alpha |v|}{\beta} <0$. If $  2 \omega -  \frac{ \alpha |v|}{\beta} > 0$ then for $t_1 > 0$ we must have that
\begin{dmath}
 \frac{\omega^2}{2} - \alpha \theta - \frac{\alpha \omega |v| }{\beta} < 0
 \label{eq:omega2condition2}
\end{dmath}
\noindent by a similar argument that yielded \eqref{eq:omega1condition2}. Similar to case $1$ above, we can show that that conditions \eqref{eq:omega2condition1} and \eqref{eq:omega2condition2} are satisfied when $H_1>0$ and $H_2 >0$. Therefore, a $C2a$ control with first phase $NP$ exists such that the state reaches $(0,0,0)$ from $q_0$, when $q_0 \in \Omega_2$.
We have characterized the initial conditions for which $C2a$ controls exist. 
\end{pf}

\begin{figure}
\centering
\includegraphics[width=0.4\textwidth]{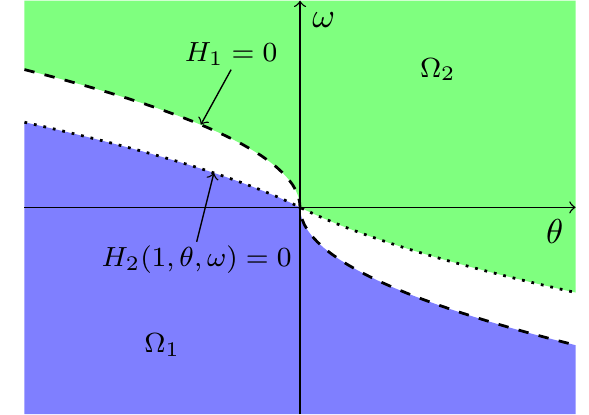}
\caption{The shaded region corresponds to the set $\Omega_1 \cup \Omega_2$ for some fixed value of $v$.}
\label{fig:Omega12}
\end{figure}

\begin{rem}
$\Omega_1 \cap \Omega_2 = \emptyset $
\end{rem}

\begin{lem}
Let $q_0 \in \Omega_3$. Then, an extremal $(q^*(t),\psi^*(t),u^*(t))$ defined on $I = [0,t_f]$ exists such that $q^*(0) = q_0$ and $q^*(t_f) = (0,0,0)$, and $u^*(t)$ is of type $C2b$.
\label{lem:omega3}
\end{lem}
\begin{pf}
We can derive the change in state due to a $C2b$ control as 
\begin{dgroup}
\begin{dmath}
v (\bar{t}_3)  = v + s_3 \beta (t_1 - t_3)
\label{eq:wnonzeroc1vt3}
\end{dmath}
\begin{dmath}
\theta(\bar{t}_3)  = \theta + \omega t_1 + \omega t_2 + \frac{1}{2} s_4 \alpha t_2^2+ (\omega + s_4 \alpha t_2) t_3
\label{eq:wnonzeroc1thetat3}
\end{dmath}
\begin{dmath}
\omega(\bar{t}_3)  =\omega + s_4 \alpha t_2
\label{eq:wnonzeroc1wt3}
\end{dmath}
\label{eq:wnonzeroc2b}
\end{dgroup}

\noindent where $s_3 \in \{1,-1\}$ and $\bar{t}_2 = t_1 + t_2$, $\bar{t}_3 = t_1+t_2+t_3$. If the second control phase is $\myNP$, then $s_4 = 1$, otherwise $s_4 = -1$. If the first control phase is $\myPP$, then $s_3 = 1$, otherwise $s_3=-1$. 


Setting $\omega(\bar{t}_3) = 0$, we can obtain
\begin{dgroup}
\begin{dmath}
t_2 = \frac{|\omega |}{\alpha}
\label{eq:wnonzeroc1t2}
\end{dmath}
\begin{dmath}
s_4 = \mysign{- \omega}
\label{eq:wnonzeroc1s2}
\end{dmath}
\label{eq:wnonzeroc1t2s2}
\end{dgroup}

Substituting \eqref{eq:wnonzeroc1t2} and \eqref{eq:wnonzeroc1s2} in \eqref{eq:wnonzeroc1thetat3} and requiring that  $\theta(\bar{t}_3) = 0$ yields
\begin{dmath}
0 = \theta + \omega t_1 + \omega  \frac{|\omega| }{\alpha} +  - \frac{\omega | \omega | }{2 \alpha} + ( 0 ) t_3
\end{dmath}
\noindent so that we can compute $t_1$ to be
\begin{dmath}
t_1 = - \frac{\omega | \omega| + 2 \alpha \theta}{2 \alpha \omega}\\
 = -\frac{H_1(\omega,\theta)}{2 \alpha \omega}
 \label{eq:t1c2b}
\end{dmath}

Note that for the control to be of type $C2b$, $t_1$ and $t_3$ must be strictly positive. In order for $t_1$ to be positive, we need $ \omega H_1(\theta,\omega) <0$. 
\begin{dmath}
\omega H_1(\theta,\omega) <0
\label{eq:omega3condition}
\end{dmath} 

Setting $v(\bar{t}_3) = 0$, we can obtain
\begin{dmath}
t_3 = \frac{v}{s_3 \beta} +  t_1
\label{eq:wnonzeroc1t3t1}
\end{dmath}

If $\frac{| v |}{\beta} < t_1 $ then then we may pick $s_3$ to be either $+1$ or $-1$ in order to ensure that $t_3 >0$. If not, then $s_3 = \mysign{v}$ yields $t_3 >0$. The only condition that is required to ensure that an extremal of the form $C2b$ exists is given in \eqref{eq:omega3condition}. These are exactly the set of points $q_0 \in \Omega_3$, proving the lemma.
\end{pf}

\begin{figure}
\centering
\includegraphics[width=0.4\textwidth]{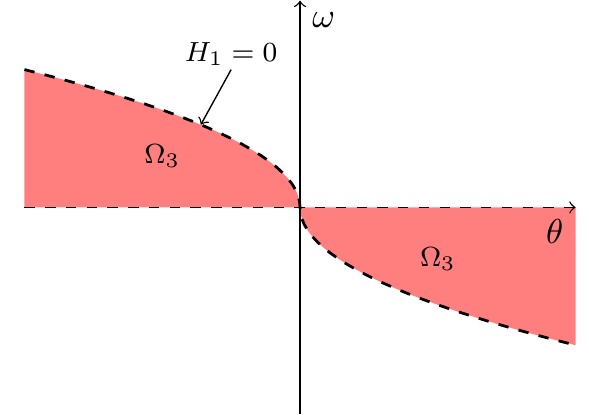}
\caption{The shaded region corresponds to the set $\Omega_3$ for some fixed value of $v$.}
\label{fig:Omega3}
\end{figure}

We have already defined sets $\Omega_1$ and $\Omega_2$, and fortunately $\Omega_1 \cap \Omega_2 = \emptyset$. This means that for initial conditions in either of those sets, we have a unique $C2a$ control which results in a transition of the state to the origin. Unfortunately, $\Omega_1 \cap \Omega_3 \neq \emptyset$ and $\Omega_2 \cap \Omega_3 \neq \emptyset$. This means that for $(v,\theta,\omega) \in \Omega_1 \cap \Omega_3 $ or $(v,\theta,\omega) \in \Omega_1 \cap \Omega_3 $, we must be able to decide whether the $C2a$ control or the $C2b$ control is faster. We will now show that the $C2a$ control is always faster.

\begin{lem}
Let $q_0 \in (\Omega_1 \cup \Omega_2 ) \cap \Omega_3$. Let the duration of the $C2a$ extremal corresponding to $q_0$ be $t_{C2a}$ and the duration of the $C2b$ extremal be $t_{C2b}$. Then
\begin{dmath}
t_{C2a} < t_{C2b}
\end{dmath}
\label{lem:C2afasterC2b}
\end{lem}
\begin{pf}
We consider the case when $(v,\theta,\omega) \in \Omega_1 \cap \Omega_3$. One can check that $\omega >0$ and $\theta <0$. The total time $t_{C2a}$ for the extremal trajectory due to a $C2a$ control is given by
\begin{dmath}
t_{C2a} = - \frac{\omega}{\alpha} + \frac{1}{\alpha} \bigpar{2 \omega^2 - 4 \alpha \theta + \alpha^2 \frac{|v|^2}{\beta^2}}^{\frac{1}{2}}
\end{dmath}

The total time $t_{C2b}$ for the extremal trajectory due to a $C2b$ control is given by
\begin{dmath}
t_{C2b} = - \frac{2 \theta}{\omega} + \frac{|v|}{\beta} 
\end{dmath}

Therefore, we can compute
\begin{dmath}
t_{C2a} - t_{C2b}= - \frac{\omega}{\alpha} + \frac{1}{\alpha} \bigpar{2 \omega^2 - 4 \alpha \theta + \alpha^2 \frac{|v|^2}{\beta^2}}^{\frac{1}{2}} +  \frac{2 \theta}{\omega} - \frac{|v|}{\beta} \\
 = \bigpar{ \frac{2 (\omega^2 - 2 \alpha \theta)}{\alpha^2} + \frac{|v|^2}{\beta^2}}^{\frac{1}{2}}  - \bigpar{\frac{\omega}{\alpha} - \frac{2 \theta}{\omega} +  \frac{|v|}{\beta} }\\
 = \bigpar{ \frac{2 \epsilon}{\alpha^2} + \frac{|v|^2}{\beta^2}}^{\frac{1}{2}}  - \bigpar{\frac{\epsilon}{\alpha \omega}+\frac{|v|}{\beta} }
\label{eq:tC2amtC2b}
\end{dmath}
\noindent where $\epsilon = \omega^2 - 2 \alpha \theta$ and $\epsilon >0$ since $(v,\theta,\omega) \in \Omega_1$. The expression above is negative, which we show by considering the difference between the term under the square root and the square of the second term in \eqref{eq:tC2amtC2b}. Note that both terms are positive. Let the difference be $\delta$, given by
\begin{dmath}
\delta = \frac{2 \epsilon}{\alpha^2} + \frac{|v|^2}{\beta^2} - \bigpar{\frac{\epsilon}{\alpha \omega}+\frac{|v|}{\beta} }^2\hfill\\
= \frac{2 \epsilon}{\alpha^2} - \frac{\epsilon^2}{\alpha^2 \omega^2} - \frac{2 \epsilon |v| }{\alpha \omega \beta}\hfill\\
= \frac{2 \epsilon}{\alpha^2 \omega^2} \bigpar{\omega^2  - \frac{\epsilon}{2}  - \frac{\alpha \omega |v| }{ \beta}}\hfill\\
= \frac{2 \epsilon}{\alpha^2 \omega^2} \bigpar{\frac{\omega^2}{2} + \alpha \theta  - \frac{\alpha \omega |v| }{ \beta}}\hfill
\end{dmath}

We have already established that $\epsilon >0$. Since $(v,\theta,\omega) \in \Omega_1 $, this implies that 
\begin{dmath}
\frac{\omega^2}{2} + \alpha \theta  + \frac{\alpha \omega |v| }{ \beta} <0\\
\hiderel{\Rightarrow} \frac{\omega^2}{2} + \alpha \theta  - \frac{\alpha \omega |v| }{ \beta} <0
\end{dmath}
\noindent since $\omega >0$. This implies that $\delta <0$, which implies that $t_{C2a} - t_{C2b} <0$. For the case of initial conditions $(v,\theta,\omega) \in \Omega_2 \cap \Omega_3$, a similar argument can be applied, which is omitted here.
\end{pf}

We have determined initial conditions for which $C2a$ or $C2b$ controls exist such that the state transitions to $(0,0,0)$. We now determine initial conditions for which $C1a$ controls exist. Recall that an extremal control is of the form $C1_{ns}$ if one of the motor torques is always $+\torqmax$ or $-\torqmax$ for the duration of the trajectory, and the other motor switches no more than twice.

\begin{lem}
Let $q_0 \in \Omega_4$. Then, an extremal $(q^*(t),\psi^*(t),u^*(t))$ defined on $I = [0,t_f]$ exists such that $q^*(0) = q_0$ and $q^*(t_f) = (0,0,0)$, and $u^*(t)$ is of the form $\bpmphase \rightarrow \apmphase \rightarrow \bpmphase $.
\label{lem:omega4bab}
\end{lem}
\begin{pf}
We first show that if $(v,\theta,\omega) \in \Omega_4$, then $\omega H_1(v,\theta,\omega) <0$ and $H_2(v,\theta,\omega) >0$.\\
 Let $\omega >0$. Then, $H_1(v,\theta,\omega) >0 \implies H_2(v,\theta,\omega) >0$. Let $\omega <0$. Then, $H_1(v,\theta,\omega) <0 \implies H_2(v,\theta,\omega)<0$. In other words, when $\omega H_1(v,\theta,\omega) >0$ then $\omega H_2 <0$ cannot be true. Thus, $(v,\theta,\omega) \in \Omega_4 \implies \omega H_1(v,\theta,\omega) <0$, $\omega H_2(v,\theta,\omega) >0$.
 
Next, we show that these conditions are sufficient to ensure a $C1_{ns}$ control with three distinct phases exists such that an initial condition $(v,\theta,\omega)$ will reach the origin.  

The equations are similar to \eqref{eq:wnonzeroc2b}, except for the change in sign of one term. However, this change is significant.
\begin{dgroup}
\begin{dmath}
v (\bar{t}_3)  = v + s_3 \beta (t_1 + t_3)
\label{eq:wnonzeroc1nsvt3}
\end{dmath}
\begin{dmath}
\theta(\bar{t}_3)  = \theta + \omega t_1 + \omega t_2 + \frac{1}{2} s_4 \alpha t_2^2+ (\omega + s_4 \alpha t_2) t_3
\label{eq:wnonzeroc1nsthetat3}
\end{dmath}
\begin{dmath}
\omega(\bar{t}_3)  =\omega + s_4 \alpha t_2
\label{eq:wnonzeroc1nswt3}
\end{dmath}
\label{eq:wnonzeroc1nsd714}
\end{dgroup}


The solutions for $t_1$ and $t_2$ remain the same as in \eqref{eq:t1c2b} and \eqref{eq:wnonzeroc1t2} respectively. The value of $s_3$ is different, which we obtain by setting $v(\bar{t}_3) = 0$:
\begin{dmath}
0 = v + s_3 \beta (t_1 + t_3)
\label{eq:c1avtfequals0}
\end{dmath}
\noindent which implies that 
\begin{dmath}
s_3 = -\mysign{v}
\end{dmath}

Now,
\begin{dmath}
t_3 = -\frac{v}{s_3 \beta} -  t_1\\
 = \frac{| v |}{\beta} -  t_1
\label{eq:wnonzeroc1nst3t1}
\end{dmath}
In order for $t_3 > 0$ we must have 
\begin{align}
 -\frac{\omega | \omega| + 2 \alpha \theta}{2 \alpha \omega} & < \frac{|v| }{\beta}\\
 \Rightarrow \frac{| \omega | }{2 \alpha } +\frac{ \theta}{ \omega} + \frac{|v| }{\beta} & > 0\\
 \Rightarrow \frac{\omega H_2(v,\theta,\omega)}{\omega^2} & > 0 \label{eq:omega4condition} 
\end{align}
From \eqref{eq:wnonzeroc1t2s2} we know that if $\omega H_1(v,\theta,\omega)  < 0$ then $t_1 > 0$. Additionally, if $\omega H_2(v,\theta,\omega) > 0$ then $t_3 > 0$. These conditions are met when $H_1(v,\theta,\omega) H_2(v,\theta,\omega) <0$. Therefore, $q_0 \in \Omega_4$ implies that a $C1_{ns}$ control exists such that the resulting trajectory reaches the origin.

%
\noindent \emph{Case 2:} For the first phase, $\torqr = -\torql$. The equations are similar to \eqref{eq:wnonzero}, again except for two changes in signs:
\begin{dgroup}
\begin{dmath}
\theta(\bar{t}_3) = \theta + \omega t_1 + \frac{s_1 \alpha t_1^2}{2}+ (\omega + s_1 \alpha t_1) t_2  + (\omega + s_1 \alpha t_1) t_3  +\frac{s_1 \alpha t_3^2}{2}
\label{eq:eqsetn1}
\end{dmath}
\begin{dmath}
\omega(\bar{t}_3) = s_1 \alpha t_1 + s_1 \alpha t_3 + \omega
\label{eq:eqsetn2}
\end{dmath}
\begin{dmath}
v(\bar{t}_3) = s_2 \beta t_2 +v
\label{eq:eqsetn3}
\end{dmath}
\label{eq:wnonzeroc1ns2}
\end{dgroup}

In order for the intervals $t_2$ and $t_3$ to be positive, we see that 
\begin{dgroup}
\begin{dmath}
s_1 ={ -\mysign{\omega}} \hiderel{\Rightarrow} t_3 \hiderel{=} \frac{| \omega | }{\alpha} - t_1 \\
\end{dmath}
\begin{dmath}
s_2 = {-\mysign{v}} \hiderel{\Rightarrow}   t_2 \hiderel{=} \frac{| v | }{\beta} 
\end{dmath}
\label{eq:proofO4c2s1s2t2} 
\end{dgroup}

Substituting for $s_1$, $t_1$ and $t_3$ in \eqref{eq:eqsetn1}, we obtain
\begin{dmath}
\frac{\alpha |v| }{| \omega | \beta } t_1 = \frac{ |\omega | }{2 \alpha} + \frac{\theta}{\omega} + \frac{ |v|}{\beta}\\
 = \frac{H_2(v,\theta,\omega)}{\omega}
 \label{eq:proofO4c2t1}
\end{dmath}
\noindent where the square terms have canceled each other out. We see that $t_1 >0$ is true when $\omega H_2 (v,\theta,\omega) >0$. For the case of $t_3$, consider
\begin{align}
t_3 &= \frac{| \omega | }{\alpha} - t_1 \\
\implies \frac{\alpha |v| }{| \omega | \beta } t_3 &= \frac{| v | }{\beta} - \frac{\alpha |v| }{| \omega | \beta } t_1\\
\implies \frac{\alpha |v| }{| \omega | \beta } t_3 &= -\bigpar{ \frac{ |\omega | }{2 \alpha} + \frac{\theta}{\omega} }
\label{eq:proofO4c2t3}
\end{align}
\noindent so that $t_3 >0$ is true when $\omega H_1 (v,\theta,\omega)<0$. Again, these conditions are met when $q_0 \in \Omega_4$.

\end{pf}

\begin{figure}
\centering
\includegraphics[width=0.4\textwidth]{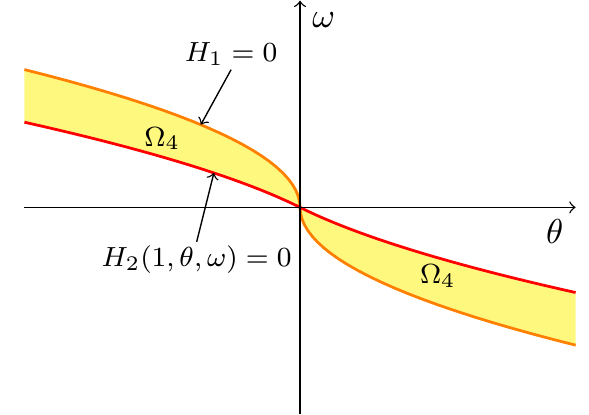}
\caption{The shaded region corresponds to the set $\Omega_4$ for some fixed value of $v$.}
\label{fig:Omega4}
\end{figure}

\begin{lem}
Let $q_0 \in \Omega_4$. Then, a unique extremal $(q^*(t),\adjointinout u^*(t))$ defined on $I = [0,t_f]$ exists such that $q^*(0) = q_0$ and $q^*(t_f) = (0,0,0)$, and $u^*(t)$ is of the form $\apmphase \rightarrow \bpmphase \rightarrow \apmphase $.
\label{lem:omega4aba}
\end{lem}
\begin{pf}
In the proof of Lemma \ref{lem:omega4bab}, we have shown that if $(v,\theta,\omega) \in \Omega_4$, then $\omega H_1(v,\theta,\omega) <0$ and $H_2(v,\theta,\omega) >0$.\\

The equations are similar to \eqref{eq:wnonzero}, again except for two changes in signs:
\begin{dgroup}
\begin{dmath}
\theta(\bar{t}_3) = \theta + \omega t_1 + \frac{s_1 \alpha t_1^2}{2}+ (\omega + s_1 \alpha t_1) t_2  + (\omega + s_1 \alpha t_1) t_3  +\frac{s_1 \alpha t_3^2}{2}
\label{eq:eqsetn1aba}
\end{dmath}
\begin{dmath}
\omega(\bar{t}_3) = s_1 \alpha t_1 + s_1 \alpha t_3 + \omega
\label{eq:eqsetn2aba}
\end{dmath}
\begin{dmath}
v(\bar{t}_3) = s_2 \beta t_2 +v
\label{eq:eqsetn3aba}
\end{dmath}
\label{eq:wnonzeroc1ns2O4aba}
\end{dgroup}

In order for the intervals $t_2$ and $t_3$ to be positive, we see that 
\begin{dgroup}
\begin{dmath}
s_1 ={ -\mysign{\omega}} \hiderel{\Rightarrow} t_3 \hiderel{=} \frac{| \omega | }{\alpha} - t_1 \\
\end{dmath}
\begin{dmath}
s_2 = {-\mysign{v}} \hiderel{\Rightarrow}   t_2 \hiderel{=} \frac{| v | }{\beta} 
\end{dmath}
\label{eq:proofO4c2s1s2t2aba} 
\end{dgroup}

Substituting for $s_1$, $t_1$ and $t_3$ in \eqref{eq:eqsetn1aba}, we obtain
\begin{dmath}
\frac{\alpha |v| }{| \omega | \beta } t_1 = \frac{ |\omega | }{2 \alpha} + \frac{\theta}{\omega} + \frac{ |v|}{\beta}\\
 = \frac{H_2(v,\theta,\omega)}{\omega}
 \label{eq:proofO4c2t1aba}
\end{dmath}
\noindent where the square terms have canceled each other out. We see that $t_1 >0$ is true when $\omega H_2 (v,\theta,\omega) >0$. For the case of $t_3$, consider
\begin{align}
t_3 &= \frac{| \omega | }{\alpha} - t_1 \\
\implies \frac{\alpha |v| }{| \omega | \beta } t_3 &= \frac{| v | }{\beta} - \frac{\alpha |v| }{| \omega | \beta } t_1\\
\implies \frac{\alpha |v| }{| \omega | \beta } t_3 &= -\bigpar{ \frac{ |\omega | }{2 \alpha} + \frac{\theta}{\omega} }
\label{eq:proofO4c2t3aba}
\end{align}
\noindent so that $t_3 >0$ is true when $\omega H_1 (v,\theta,\omega)<0$. Again, these conditions are met when $q_0 \in \Omega_4$.

\end{pf}

\begin{lem}
Let $q_0 \in S_5$. Then, an extremal $(q^*(t),\psi^*(t),u^*(t))$ defined on $I = [0,t_f]$ exists such that $q^*(0) = q_0$ and $q^*(t_f) = (0,0,0)$, and $u^*(t)$ is of the form $\apmphase \rightarrow \bpmphase $.
\label{lem:omega5}
\end{lem}
\begin{pf}
Since $H_1 = 0$  and $H_2 \neq 0$ then $\omega v \neq 0$. Consider the equations \eqref{eq:wnonzeroc1ns2} wherein $t_3 = 0$. The solutions for $t_1$ and $t_2$ are given by 
\begin{dgroup}
\begin{dmath}
t_1 = \frac{|\omega | }{\alpha}
\label{eq:H1orH20sol1}
\end{dmath}
\begin{dmath}
t_2 = \frac{|v | }{\beta}
\label{eq:H1orH20sol2}
\end{dmath}
\end{dgroup}

Clearly $t_1$ and $t_2$ are positive precisely when $\omega v \neq 0$. Thus, $q \in S_5$ implies that a control of the form above exists such that the resulting trajectory reaches the origin.
\end{pf}

\begin{lem}
Let $q_0 \in S_{6}$. Then, an extremal $(q^*(t),\psi^*(t),u^*(t))$ defined on $I = [0,t_f]$ exists such that $q^*(0) = q_0$ and $q^*(t_f) = (0,0,0)$, and $u^*(t)$ is of the form $\bpmphase \rightarrow  \apmphase$.%
\label{lem:omega6}
\end{lem}
\begin{pf}
Since $H_2 = 0$  and $H_1 \neq 0$ then again we can conclude that $\omega v \neq 0$. The equations \eqref{eq:wnonzeroc1nsd714} wherein $t_3 = 0$ have solutions
\begin{dgroup}
\begin{dmath}
t_1 = \frac{|v | }{\beta}
\end{dmath}
\begin{dmath}
t_2 = \frac{|\omega | }{\alpha}
\end{dmath}
\end{dgroup} Again, $t_1$ and $t_2$ are positive precisely when $\omega v \neq 0$.Thus, $q \in S_{6}$ implies that a control of the form above exists such that the trajectory reaches the origin.
\end{pf}

\begin{lem}
Let $q_0 \in L_v$. Then, an extremal $(q^*(t),\psi^*(t),u^*(t))$ defined on $I = [0,t_f]$ exists such that $q^*(0) = q_0$ and $q^*(t_f) = (0,0,0)$, and $u^*(t)$ is of the form $\bpmphase$.%
\label{lem:Lv}
\end{lem}
\begin{pf}
Since $q_0 \in L_v$, $\omega = \theta =0$ and $v \neq 0$. Substituting $\omega = \theta =0$ in \eqref{eq:wnonzeroc1nsd714} we immediately see that $t_2 = t_3 = 0$ and $t_1 = \frac{| v | }{\beta}$. Thus a control of the form $\bpmphase$ exists such that the trajectory from $q_0$ reaches the origin.
\end{pf}
\begin{lem}
Let $q_0 \in L_\omega$. Then, an extremal $(q^*(t),\psi^*(t),u^*(t))$ defined on $I = [0,t_f]$ exists such that $q^*(0) = q_0$ and $q^*(t_f) = (0,0,0)$, and $u^*(t)$ is of the form $\apmphase$.%
\label{lem:Lomega}
\end{lem}
\begin{pf}
Since $q_0 \in L_\omega$, $H_1(v,\theta,\omega) = H_2(v,\theta,\omega) = v =0$.  From \eqref{eq:proofO4c2t3} we see that $t_3 = 0$ and from \eqref{eq:proofO4c2s1s2t2} we see and $t_2 = 0$. Further, $t_3 = 0$ implies that $t_1 = \frac{| \omega | }{\alpha}$, due to \eqref{eq:proofO4c2s1s2t2}. Thus,  a control of the form $\apmphase$ exists such that the trajectory from $q_0$ reaches the origin.
\end{pf}

\begin{lem}
Let  $q_0 = (v_0,\theta_0,\omega_0) \in \R^3$. Let there exist an extremal $(q^*(t),\psi^*(t),u^*(t))$ defined on $I = [0,t_f]$ such that $q^*(0) = q_0$ and $q^*(t_f) = (0,0,0)$, such that $u^*(t)$ is a singular control. Then there exists a $C1_{ns}$ control $\bar{u}^*(t)$ and corresponding extremal $(\bar{q}^*(t),\bar{\psi}^*(t),\bar{u}^*(t))$ defined on $[0,\bar{t}_f]$ such that $\bar{q}^*(0) = q_0$, $\bar{q}^*(\bar{t}_f) = (0,0,0)$ and $\bar{t}_f = t_f$.
\label{lem:ignoreC1s}
\end{lem}
\newcommand{\cpdr}{\omega_r}
\newcommand{\cpdl}{\omega_l}
\newcommand{\cpddr}{\dot{\omega}_r}
\newcommand{\cpddl}{\dot{\omega}_l}

\begin{pf}
Consider the quantities $\omega_r$ and $\omega_l$ given by
\begin{dmath}
\bmat{\cpdr \\ \cpdl} =\bmat{c_1 & c_2 \\ c_2 & c_1} \bmat{\pdr \\ \pdl} 
\end{dmath}
\noindent where $c_1$ and $c_2$ are the parameters in \eqref{eq:wheelspeeddynamics}. The dynamics of these quatities are then simply
\begin{dmath}
\bmat{\cpddr \\ \cpddl} = \bmat{\torqr \\ \torql} 
\end{dmath}

Note that $c_1 \neq c_2$ implies that  $\cpdr = \cpdl = 0 \iff \pdr = \pdl = 0$. At the goal state, $v = \omega =0$ which implies that $\cpdr = \cpdl  = 0$. 
Since one motor never switches, the duration of any $C1$ extremal control is  exactly the time taken to bring the larger (in magnitude) of $\cpdr  $ and $\cpdl $  to zero. Thus,
\begin{dmath}
t_f = \frac{  \mathrm{max}\bigpar{ |  \cpdr(0) | ,| \cpdl(0) | }}{\torqmax}\\
= \frac{  \mathrm{max}\bigpar{ |  c_1 \pdr(0) + c_2 \pdl(0)  | ,|  c_2 \pdr(0) + c_1 \pdl(0) | }}{\torqmax}
\end{dmath}
Any $C1$ extremal control with initial condition $q_0$ must have this same duration. Let $| \cpdr(0) | > | \cpdl(0)|$. The right motor torque $\torqr^*(t)$ is given by
\begin{dmath*}
\torqr^*(t) = -s \torqmax
\end{dmath*}
\noindent where $s = \mysign{\cpdr(0)}$.

The singular control $u^*(t)$ is such that $v(t_f) = \omega(t_f) = \theta(t_f) = \cpdl(t_f) = \cpdr(t_f) = 0$. We can compute $\cpdl(t_f)$ as 
\begin{dmath*}
{\cpdl(t_f) = \gamma(\torql(t)) = \cpdl(0) + \int_{0}^{\bar{t}_3}\torql(\tau) d\tau}
\end{dmath*}
\noindent where $\gamma $ is a functional with argument $\torql(t)$. We can compute $\theta(t_f)$ as 
\begin{dmath*}
\theta(t_f) = \theta_0 + \omega_0 \bar{t}_3 + \int_{0}^{\bar{t}_3} \int_{0}^{t} \frac{r}{2 b}( \pddr(\tau) - \pddl(\tau)) d\tau\\
= \theta_0 + \omega_0 \bar{t}_3 + \int_{0}^{\bar{t}_3} \int_{0}^{t} \frac{r}{2 b(c_1 - c_2 )} ( \torqr(\tau) - \torql(\tau)) d\tau\\
= \theta_0 + \omega_0 \bar{t}_3 - c_3 \frac{s \torqmax \bar{t}_3^2 }{2}  - c_3 \int_{0}^{\bar{t}_3} \int_{0}^{t} \torql(\tau) d\tau dt \\
= k(q_0) - c_3 h(\torql(t))
\end{dmath*}
\noindent where 
\begin{dgroup}
\begin{dmath}
c_3 = \frac{r}{2 b(c_1 - c_2 )}
\end{dmath}
\begin{dmath}
k(q_0) = \bigpar{ \theta_0+ \omega_0 \bar{t}_3 -  c_3 \frac{s \torqmax \bar{t}_3^2 }{2}}
\end{dmath}
\begin{dmath}
h(\torql(t)) = \int_{0}^{t_f} \int_{0}^{t} \torql(\tau) d\tau dt
\end{dmath}
\end{dgroup}

Note that $h(\torql(t))$ is also a functional. The control $u^*(t)$ can be arbitrary, as long as $\gamma^*(\torql(t)) = 0$ and  $h(\torql^*(t)) = k(q_0 ) / c_3$. We claim that a control $\torql(t)$  exists such that  $\torql(t)$ only switches twice and also satisfies these conditions. Consider the control $\torql^{ns}(t)$ defined as
\begin{dmath}
\torql^{ns}(t) =
\begin{cases}
-s \torqmax & \mbox{ if } t < t_1\\
 s \torqmax & \mbox{ if } t_1 < t < t_1 + t_2\\
-s \torqmax & \mbox{ if } t_1 + t_2 < t < \bar{t}_3
\end{cases}
\end{dmath}

Note that this means $u^{ns}(t)$ is of the form $\bpmphase \rightarrow \apmphase \rightarrow \bpmphase$, with three phases of duration $t_1$, $t_2$ and $t_3$. If the goal state is the origin, then $t_2= \frac{| \omega | }{\alpha}$ and $t_1 + t_3 = \frac{| v | }{\beta}$. Thus, $h(\torql^{ns}(t))$ depends only on $t_1$.
%

Consider the function $g(t_1)$ given by
\begin{dmath}
g(t_1) = h(\torql^{ns}(t)) \hfill\\
 = \frac{s \torqmax}{2} (t_2^2 -2 t_1 t_2 - t_1^2) -   \frac{s \torqmax}{2} (\bar{t}_3 - (t_1+t_2))^2 +  s \torqmax (t_2 - t_1)  (\bar{t}_3 - (t_1+t_2))
\end{dmath}
\noindent where $t_1 \in [0,(\bar{t}_3 - t_2)]$ is the only unknown. In order for $\torql^{ns}(t)$ to be a valid control, we must have $g(t^{ns}) = h(\torql^{ns}(t)) = k(q_0) = h(\torql^{*}(t))$ for some $t^{ns} \in [0,(\bar{t}_3 - t_2)]$.

We can compute the derivative of $g$ with respect to $t_1$ and obtain 
\begin{dmath}
\frac{d}{dt} g(t_1) = -2 t_2 
\end{dmath}
\noindent which implies that $g(t_1)$ is a strictly decreasing function of $t_1$. Since $\torql^*(t)$ is bounded and $\int_{0}^{\bar{t}_3}\torql^*(\tau) d\tau = -\cpdl(0)$, $h(\torql^*(t))$ is bounded. One can show that $ g(\bar{t}_3-t_2) \leq h(\torql^{*}(t)) \leq g(0) $. Thus, we can find $t^{ns} \in [0, \bar{t}_3-t_2]$ such that $g(t^{ns}) = h(\torql^{ns}(t)) = k(q_0) / c_3 $. Due to the selection of $t_2$, $\gamma(\torql^{ns}(t)) = 0$. Thus, $\torql^{ns}(t)$ is a valid extremal control such that the state at $t = t_f$ reaches the origin from $q_0$, proving the result.

\end{pf} 
\begin{rem}
Lemma \ref{lem:ignoreC1s} implies that if we consider all possible $C1_{ns}$ trajectories from an initial state $q_0$, then we do not need to consider singular controls even if they exist, since one of the $C1_{ns}$ controls will result in the trajectory reaching the goal state in the same time.
\end{rem}

\begin{lem}
Let $q_0 \in \Omega_3 \cap ( \Omega_4 \cup S_5 \cup S_{6} \cup L_v \cup L_\omega )$  . Let the duration of the $C2b$ extremal corresponding to $q_0$ be $t_{C2b}$ and the duration of the $C1_{ns}$ extremal be $t_{C1ns}$. Then
\begin{dmath}
t_{C1ns} \leq t_{C2b}
\end{dmath}
\label{lem:C2bslow}
\end{lem}
\begin{pf}
The duration $t_{C1ns}$ of any $C1_{ns}$ control and initial condition $(v,\theta,\omega)$ is given by
\begin{dmath}
t_{C1ns} = (t_1 + t_3) + t_2\\
 = \frac{|v|}{\beta}+\frac{|\omega|}{\alpha}
\end{dmath} 
\noindent which is independent of $\theta$.

A $C2b$ control from $(v,\omega,\theta)$ has total duration 
\begin{dmath}
t_{C2b} = (t_1 + t_3) + t_2\\
 = 2 t_1 + \frac{v}{s_3 \beta} + \frac{|\omega |}{\alpha}
\end{dmath}
\noindent where we used \eqref{eq:wnonzeroc1vt3} and \eqref{eq:wnonzeroc1t2} with $v(\bar{t}_3) = 0$. If $t_1 \leq \frac{| v | }{\beta}$ then we must pick $s_3 = \mysign{v}$ (see the last paragraph of the proof of Lemma \ref{lem:omega3}) so that 
\begin{dmath}
t_{C2b} = 2 t_1 + \frac{|v|}{\beta} + \frac{|\omega|}{\alpha}\\
 > \frac{|v|}{\beta} + \frac{|\omega|}{\alpha} 
\end{dmath} 
\noindent since $t_1 > 0$. If $t_1 > \frac{| v | }{\beta}$ then 
\begin{dmath}
t_{C2b} = 2 t_1 + \frac{v}{s_3 \beta} + \frac{|\omega|}{\alpha}\hfill\\
> 2 t_1 - \frac{|v|}{\beta} + \frac{|\omega|}{\alpha}\hfill\\
> t_1 +  t_1 - \frac{|v|}{\beta} + \frac{|\omega|}{\alpha}\hfill\\
> t_1 + \frac{|\omega|}{\alpha}\hfill\\
> \frac{|v|}{\beta} + \frac{|\omega|}{\alpha}    \hfill
\end{dmath}
Thus, whenever both $C1_{ns}$ and $C2b$ controls exist, the $C1_{ns}$ control is always faster.    
\end{pf}

\section{Regular Synthesis}
\label{sec:regularsynthesis}
Given a point $q \in \mathbb{R}^3$, we can determine the form of the time-optimal control $u_{q}^*(t)$ defined on $[0,t_f]$ which results in a trajectory $q^*(t)$ corresponding to a minimum-time transition from $q$ to the origin. The trajectory $q^*(t)$ is the solution to a differential equation driven by the discontinuous input signal $u_q^*(t)$.

For each $t \in [0,t_f]$, $u^q*(t) \in \{\myPN,\myNP,\myPP,\myNN \}$. One can immediately define a feedback control $v^*(q) = u_{q}^{*}(0)$. The resulting closed-loop system under $v^*(q)$ is now a differential equation with a discontinuous right-hand side. It is not necessary that the solutions of this new dynamical system correspond to $q^*(t)$. 

In order to prove that our feedback law results in time-optimal behaviour, we must construct an optimal regular synthesis~\cite{Piccoli00} and show that the solutions of the closed-loop dynamical system under the feedback law only produces the trajectories defined by the optimal regular synthesis.

We begin by introducing appropriate definitions. 
Let the set of admissible controlled trajectories that result in a transition from a state $q$ to the origin be denoted by $\admissct$. Thus, elements of $\admissct$ are pairs $(\gamma,\eta)$ where $\eta$ is an admissible control defined on some interval $[0,t_f]$ and $\gamma$ is the resulting trajectory. Moreover, we denote the initial condition of $\gamma$ as $\gamma^{-}$.
\begin{defn}
A presynthesis for $\mc{P}$ is a subset $\Gamma$ of $\admissct$ such that 
\begin{enumerate}
 \item[PS] Whenever $(\gamma_1, \eta_1) \in \Gamma$, $(\gamma_2, \eta_2) \in \Gamma$ and $\gamma_1^-  = \gamma_2^-$, it follows that $(\gamma_1, \eta_1) = (\gamma_2, \eta_2) $
\end{enumerate}
 
\end{defn}

This definition merely stipulates that a presynthesis must assign a unique trajectory for each initial condition. \def\mydom{\mathrm{Dom}}
The set $\mydom(\Gamma) = \{\gamma^- \colon (\gamma, \eta) \in \Gamma \}$ is called the domain of $\Gamma$. In other words, it is the set of initial conditions $x \in \R^3$ for which a controlled trajectory exists in $\Gamma$ which results in a transition from $x$ to the origin. We say that $\Gamma$ is a presynthesis
on a set $S$ if $\Gamma$ is a presynthesis and $S = \mydom(\Gamma)$. 

\begin{defn} If $\Gamma$ is a presynthesis such that $\mydom(\Gamma)$ consists of all points that can be steered to the origin by means of a pair belonging to $\admissct$, then we say that $\Gamma$ is total.
\end{defn}

Thus, a total presynthesis is one which does not leave out initial conditions $x$ for which some admissible trajectory in $\admissct$ could have resulted in a transition from $x$ to the origin. Given a presynthesis $\Gamma$ and a point $x \in \mydom(\Gamma)$, we will always use $(\gamma_x, \eta_x)$ to denote the unique controlled trajectory $(\gamma, \eta) \in \Gamma$ such that $\gamma^{-} = x$.

\begin{defn}
A presynthesis on a set $S$ is memoryless if whenever $x \in S$ and $t \in \mydom( \eta_x)$ it follows that $y = \gamma_x(t)$ belongs to $S$ and $ \eta_y$ is the restriction of $\eta_x$ to the interval $[t,t_f]$. A synthesis is a memoryless presynthesis.
\end{defn}

\begin{defn}
 If each pair of a presynthesis $\Gamma$ is optimal (resp., extremal), then we say that $\Gamma$ is optimal (resp., extremal)
\end{defn}

We can construct two different optimal regular syntheses $\Gamma_1$ and $\Gamma_2$. The difference between them is seen in the control for points in the set $\Omega_4$. 
  \begin{prop}
  For every $x \in (\Omega_1 \cup \Omega_2)$, $\Omega_4$, $S_5$, $S_{6}$, $L_v$, or $L_\omega$, let $(q_x^*(t),u^*_x(t))$ be the unique extremal defined for $x$ by Lemma \ref{lem:omega12}, \ref{lem:omega4aba}, \ref{lem:omega5}, \ref{lem:omega6}, \ref{lem:Lv}, or \ref{lem:Lomega} respectively. Let $\Gamma_1 = \cup_{x \in \mathbb{R}^3} (q_x^*(t),u^*_x(t))$. Then, $\Gamma_1$ defines an optimal regular synthesis for the time-optimal control problem.
   \label{prop:regsynth1}
 \end{prop}

  \begin{pf}
The set $(\Omega_1 \cup \Omega_2) \cup \Omega_4 \cup S_5 \cup S_{6} \cup L_v \cup L_\omega \cup \{ 0 \}$ is equal to $\R^3$. For each $x \in \R^3$, the controlled trajectorys $(q_x^*(t),u^*_x(t))$ is unique and extremal, as shown in the appropriate Lemma mentioned in the proposition. Thus, $\Gamma_1$ forms a total extremal presynthesis. It is straightforward to check that $\Gamma_1$ is memoryless. Thus, $\Gamma_1$ is a total extremal synthesis. 

Next, we show that $\Gamma_1$ is a regular synthesis (Definition~$2.12$,~\cite{Piccoli00}). The conditions used to define a regular synthesis rely on numerous other definitions. Any term appearing in the rest of this proof which has not been defined in this paper has been defined in~\cite{Piccoli00}. We refer the reader to~\cite{Piccoli00} for these definitions.

To show that a synthesis is regular, we must show that a certain cost function satisfies weak continuity conditions (Definition $2.8$, \cite{Piccoli00}) and that $\Gamma_1$ is $(f,L)$-differentiable (Definition $2.9$, \cite{Piccoli00}) at all points in $\mydom(\Gamma_1)$ excluding a thin set (Definition $2.10$,~\cite{Piccoli00}).

The cost function $V_{\Gamma} \colon \R^3 \to \R$ is simply the time taken to transition from a given initial condition to the origin. The Function is continuous in $\R^3$, which can be seen through the analytical expressions obtained. Thus, $V_{\Gamma}$ satisfies the conditions of Definition $2.8$ in~\cite{Piccoli00}. 

The property of $(f,L)$-differentiability is more complicated to show. The Lagrangian function $L \colon \R^3 \times U \to \R$ common in optimal control problems reduces to the constant function $L(q,u) = 1$ in the case of time-optimal control for reaching a single goal state. The dynamics $f(q,u) = A q + B u$ is linear. Define $\tilde{f}(q,u) = [f(q,u)^T L(q,u)]^T$. If a control $\eta(t)$ is given, then $\tilde{f}_{\eta}(q,t) = \tilde{f}(q,\eta(t))$. Then,
\begin{dmath}
 \tilde{f}_{\eta}(q,t) = \bmat{A q + B \eta(t) \\ 1}
\end{dmath}
We define the function $\rho_{\tilde{f},\Gamma_1,\bar{v},t_f}(v)$ as in~\cite{Piccoli00}, and compute it as
 \begin{dmath}
\rho_{\tilde{f},\Gamma_1,\bar{v},t_f}(v)=  \tilde{f}_{\eta_{\bar{x}} + v}(\gamma_{\bar{x}}(t) ,t) -\tilde{f}_{\eta_{\bar{x} }} (\gamma_{\bar{x}}(t) ,t) \hfill\\
  = \bmat{A \gamma_{\bar{x}}(t) ,t) + B \eta_{\bar{x}}(t) + B v - A \gamma_{\bar{x}}(t) ,t) - B \eta_{\bar{x}}(t) \\ 1-1} \hfill\\
  = \bmat{B v\\0}\hfill
  \label{eq:fldiffres1}
 \end{dmath}
 
 The set $S_{thin} = \{ q \in \R^3 \colon q \in S_5 \cup S_{6} \cup L_v \cup L_{\omega} \cup \{0\} \} = \R^3 \backslash (\Omega_1  \cup \Omega_2 \cup \Omega_4) $ is a thin set based on definition $2.10$ in~\cite{Piccoli00}, where the only measure-zero set is the singleton containing the origin. Because the extremal controls are constant on each of the sets $\Omega_1$, $\Omega_2$ and $\Omega_4$, we have that any extremal control is piece-wise constant, with no more than two points of discontinuity. Thus,
 \begin{dmath}
{ D \tilde{f}_\eta(q,t) =  \bmat{A & 0 \\ 0 &0 } \mbox{ if } q \in \R^3 \backslash S_{thin} }
\label{eq:Dftilde}
 \end{dmath}

We compute the following norm for points $y \in \R^3 \backslash S_{thin}$
 \begin{dmath}
  \left\|  \tilde{f}_{\eta_{x}}(y ,t) - \tilde{f}_{\eta_{x}}(\gamma_{\bar{x}}(t) ,t) - D \tilde{f}_{\bar{x}}(\gamma_{\bar{x}}(t) ,t) (y - \gamma_{\bar{x}}(t)) \right\| \\
  = \left\| \bmat{A y + B \eta_x(t) - A \gamma_{\bar{x}}(t)  - B \eta_x(t)\\1 - 1 } - D \tilde{f}_{\bar{x}}(\gamma_{\bar{x}}(t) ,t) (y - \gamma_{\bar{x}}(t))  \right\| \\ 
  = \left\| \bmat{ A (y - \gamma_{\bar{x}}(t)) \\ 0 }   - \bmat{A & 0 \\ 0 &0 } (y - \gamma_{\bar{x}}(t)) \right\|
  =0
  \label{eq:fldiffres2}
 \end{dmath}
\noindent where we have used \eqref{eq:Dftilde}. 
The right hand sides of \eqref{eq:fldiffres1} and \eqref{eq:fldiffres2} immediately show that the conditions $DC1$ and $DC2$ of Definition $2.9$ in~\cite{Piccoli00} are satisfied by $f$ and $L$, for points $q \in \R^3 \backslash S_{thin}$. Thus, $\Gamma_1$ is $(f,L)$-differentable at a point $\bar{x} \in \R^3 \backslash S_{thin}$. Thus, based on Definition $2.12$ in~\cite{Piccoli00}, $\Gamma_1$ is regular, where the thin set is $S_{thin}$ defined above. 

We have established that $\Gamma_1$ is a total extremal regular synthesis. Since we are concerned with the time-optimal control problem with a single goal state, we have that $V_{\Gamma_1} (0) = 0$, and so condition $(2.11)$ in~\cite{Piccoli00} is satsified. By condition (b) of Theorem $2.13$ in~\cite{Piccoli00}, we can conlude that $\Gamma_1$ is an optimal regular synthesis.
  \end{pf}

\begin{prop}
For every $x \in (\Omega_1 \cup \Omega_2)$, $\Omega_4$, $S_5$, $S_{6}$, $L_v$, or $L_\omega$, let $(q_x^*(t),u^*_x(t))$ be the unique extremal defined for $x$ by Lemma \ref{lem:omega12}, \ref{lem:omega4bab}, \ref{lem:omega5}, \ref{lem:omega6}, \ref{lem:Lv}, or \ref{lem:Lomega} respectively. Let $\Gamma_2 = \cup_{x \in \mathbb{R}^3} (q_x^*(t),u^*_x(t))$. Then, $\Gamma_2$ defines an optimal regular synthesis for the time-optimal control problem.
\end{prop}
\begin{pf}
The arguments are similar to those in the proof of Proposition \ref{prop:regsynth1} and therefore the proof is omitted.
 \end{pf}

Lemmas \ref{lem:omega12}-\ref{lem:C2bslow} have been used to construct two distinct optimal regular synthesis. We now define two control laws, one corresponding to each of these syntheses.

Consider the following (sub)sets:
\begin{align}
\Omega_{4}^{v+} &= \{ q\in \Omega_{4} : v >  0 \} \\
\Omega_{4}^{v-} &= \{ q\in \Omega_{4} : v < 0 \} \\
 \Omega_{4}^{\omega+} &= \{ q\in \Omega_{4} : \omega >  0 \} \\
 \Omega_{4}^{\omega-} &= \{ q\in \Omega_{4} : \omega < 0 \} \\
 S_{5}^{\omega+} &= \{ q\in S_{5} : \omega >  0 \} \\
 S_{5}^{\omega-} &= \{ q\in S_{5} : \omega < 0 \} \\
 S_{6}^{v+} &= \{ q\in S_{6} : v >  0 \} \\
 S_{6}^{v-} &= \{ q\in S_{6} : v < 0 \} \\
 L_{v}^{+} &= \{ q\in L_v : v >  0 \} \\
 L_{v}^{-} &= \{ q\in L_v : v < 0 \} \\
 L_{\omega}^{+} &= \{ q\in L_{\omega} : \omega >  0 \} \\
 L_{\omega}^{-} &= \{ q\in L_{\omega} : \omega < 0 \} 
 \label{eq:subsetsforfb}
\end{align}

The feedback law corresponding to $\Gamma_1$ is 

\begin{dmath}
u_{fb1}(q) = 
\begin{cases}
(+\torqmax,+\torqmax) &\mbox{ if } q \in \Omega_4^{v-} \cup L_v^{-} \cup S_{6}^{v-}  \\
(+\torqmax,-\torqmax) &\mbox{ if } q \in \Omega_2 \cup L_{\omega}^{-} \cup S_5^{\omega-} \\
(-\torqmax,+\torqmax) &\mbox{ if } q \in \Omega_1 \cup L_{\omega}^{+} \cup S_5^{\omega+}\\
(-\torqmax,-\torqmax) &\mbox{ if } q \in \Omega_4^{v+} \cup L_v^{+} \cup S_{6}^{v+} \\
(0,0) &\mbox{ if } q=(0,0,0)
\end{cases}
\label{eq:statebasedcontrol1}
\end{dmath}

The feedback law corresponding to $\Gamma_2$ is 
\begin{dmath}
u_{fb2}(q) = 
\begin{cases}
(+\torqmax,+\torqmax) &\mbox{ if } q \in  L_v^{-} \cup S_{6}^{v-}  \\
(+\torqmax,-\torqmax) &\mbox{ if } q \in \Omega_2 \cup \Omega_4^{\omega-} \cup L_{\omega}^{-} \cup S_5^{\omega-} \\
(-\torqmax,+\torqmax) &\mbox{ if } q \in \Omega_1 \cup \Omega_4^{\omega+} \cup L_{\omega}^{+} \cup S_5^{\omega+}\\
(-\torqmax,-\torqmax) &\mbox{ if } q \in  L_v^{+} \cup S_{6}^{v+} \\
(0,0) &\mbox{ if } q=(0,0,0)
\end{cases}
\label{eq:statebasedcontrol2}
\end{dmath}

Controls \eqref{eq:statebasedcontrol1} and \eqref{eq:statebasedcontrol2} differ when $(v,\theta,\omega) \in \Omega_4$. The first one is such that the resulting trajectory intersects the surface $H_1(v,\theta,\omega) = 0$ and the second one has a resulting trajectory which intersects the surface $H_2(v,\theta,\omega)$.

Note that the closed loop system can be viewed as a continuous system with a discontinuous control input. This results in a right-hand side which is continuous except for a measure-zero set $M$. For such systems, one can define solutions in multiple ways~\cite{Cortes08}, including Filippov and Caratheodory solutions. In order to take advantage of a right-uniqueness theorem in~\cite{Filippov88}, we utilize definition a) in $\S$4. We define the set-valued map $F(t,q)$ for each $t \in \R $ and $q \in \R^3$ as the smallest convex closed set containing the limit values of the vector valued function $f(t,q^*)$  for $(t,q^*) \notin M$, $q^* \rightarrow q$, and constant $t$. A solution of the closed loop system is defined to be a solution of the differential inclusion 
\begin{dmath}
 \dot{q} \in F(t,q)
 \label{eq:diffincl}
\end{dmath}

Furthermore, we are concerned with the notion of right-uniqueness of the solutions of the closed-loop system (see 1, $\S$10 in~\cite{Filippov88}). For equation $\dot{q} = f(t,q)$, right uniqueness holds at a point $(t_0,q_0)$ if there exists $t_1 > t_0$ such that each two solutions $q(t)$ of this equation satsifying $q(t_0) = q_0$ coincide on the interval $[t_0,t_1]$ or on the part of the interval on which both solutions are defined.

\begin{lem}
Consider the feedback law \eqref{eq:statebasedcontrol1} for the system \eqref{eq:dynamicalsystem}. For every initial condition $q_0 \in \mathbb{R}^3$, the solutions of the closed-loop system corresponds to the unique controlled trajectory in $\Gamma_1$ corresponding to $q_0$.
\label{lem:prooffb1}
\end{lem}
\begin{pf}
The feedback system $\dot{q} = f(q,u)$ is converted into the closed loop system 
\begin{equation}
\dot{q} = g(q)
\label{eq:closedloop1}
\end{equation}
\noindent by use of feedback \eqref{eq:statebasedcontrol1}. The vector field $g$ is discontinuous on the surfaces $H_1(q) = 0$ and $H_2(q) = 0$. It is easy to show that the unique extremal solution for any initial condition $q_0 \in \R^3$ is a solution of the closed loop system $\dot{q} = g(q)$ based on the differential inclusion \eqref{eq:diffincl}. If we show that the solutions of \eqref{eq:closedloop1} are unique for any initial condition, then we have proved the lemma.

The right uniqueness of the solutions of \eqref{eq:closedloop1} can be determined based on Theorems 2 and 4, $\S$10 in ~\cite{Filippov88}. Theorem 2 provides conditions which determine when the solutions to a system $\dot{q} = f(t,q)$ defined on a domain $G$ (where $f$ is discontinouous on a surface $S$ of codimension $1$) are (right) unique. Theorem 4 provides conditions which guarantee that solutions evolving along the intersection of multiple such surfaces are unique.

In order to apply Theorem 4, $\S$10 in ~\cite{Filippov88}, we must partition $\R^3$ appropriately (see Appendix \ref{app:filippv}) and show that the vector fields defined on these partitions meet certain conditions. Furthermore, the solutions of the discontinuous system must be, loosely speaking, compatible with this partition. The partition is based on the surfaces $H_1(q)=0$ and $H_2(q) = 0$ as follows. These surfaces divide $\R^3$ into six regions, and intersect along the two lines $L_v$ and $L_\omega$. In turn, these two lines intersect at the origin, and divide the suraces into four regions.

First, consider the following subsets of $S_5$:
\begin{align}
S_5^{++} &= \{ q \in \R^3 \colon H_1(q) = 0, \omega>0, v >0\}\\
S_5^{+-} &= \{ q \in \R^3 \colon H_1(q) = 0, \omega>0, v <0\}\\
S_5^{-+} &= \{ q \in \R^3 \colon H_1(q) = 0, \omega<0, v >0\}\\
S_5^{--} &= \{ q \in \R^3 \colon H_1(q) = 0, \omega<0, v <0\}
\end{align}
\noindent and the following subsets of $S_{6}$:
\begin{align}
S_{6}^{++} &= \{ q \in \R^3 \colon H_2(q) = 0, \omega>0, v >0\}\\
S_{6}^{+-} &= \{ q \in \R^3 \colon H_2(q) = 0, \omega>0, v <0\}\\
S_{6}^{-+} &= \{ q \in \R^3 \colon H_2(q) = 0, \omega<0, v >0\}\\
S_{6}^{--} &= \{ q \in \R^3 \colon H_2(q) = 0, \omega<0, v <0\}\\
\end{align}

Condition 1) is satisfied immediately, since the solutions, which are extremals, have only two points of switching. Condition 2) is met based on applying Theorem 2, $\S$10, \cite{Filippov88} to the hypersurfaces . Condition 3) is satsifed by proper construction of the hypersurfaces $S_i^k$.
\end{pf}

\begin{lem}
Consider the feedback law \eqref{eq:statebasedcontrol2} for the system \eqref{eq:dynamicalsystem}. For every initial condition $q_0 \in \mathbb{R}^3$, the solutions of the closed-loop system corresponds to the unique controlled trajectory in $\Gamma_2$ corresponding to $q_0$.
\end{lem}

\begin{rem}
The case when $\omega_d \neq 0$ is treated in the appendix. The total durations of all valid extremals resulting in a transition from any $q_0$ to any $q_d$ have been derived, along with the motor switching times. No feedback law is developed, unlike the case when $\omega_d = 0$.
\end{rem}

\subsection{Simulations}
\begin{figure}[tb]
\centering
\includegraphics[width=0.4\textwidth]{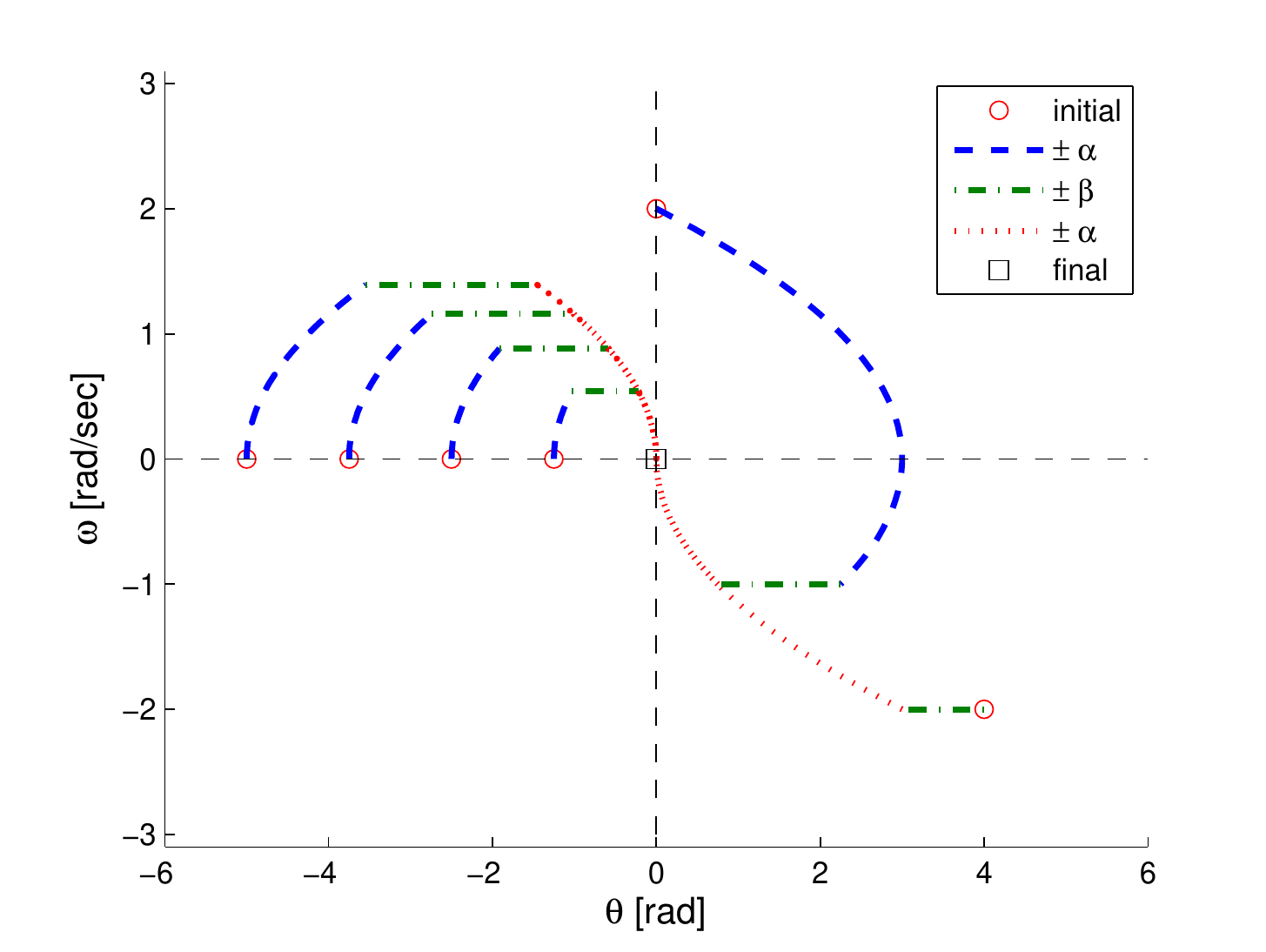}
\caption{Time-optimal trajectories for different initial conditions plotted in the $\theta-\omega$ plane. For all simulations, $v(0)=1 m/s$. The initial and final values of $(\theta,\omega)$ are marked by circles and squares respectively. Trajectories corresponding to control phases $\bpmphase$ are represented by dashed-dotted  lines and those for $\apmphase$ are represented by dashed or dotted lines.}
\label{fig:phase}
\end{figure}

For any initial condition, we can compute the time-optimal control using the method above, and simulate the open-loop implementation of this control. The results for six initial conditions are plotted in the plane $v = 0$ in Figure~\ref{fig:phase}. For all plotted trajectories, $v(0) = 1 m/s$. The circle indicates the initial values of $\theta$ and $\omega$ for each trajectory. The time-optimal control for the for the initial condition $(1 m/s,4 rad,-2 rad/s)$ is a $C1_{ns}$ control. The time-optimal controls for the remaining initial conditions are $C2$ controls. The open-loop controls result in all trajectories reaching the origin, as can be seen in Figure~\ref{fig:phase}. 

\begin{figure}[tb]
\centering
\includegraphics[width=0.4\textwidth]{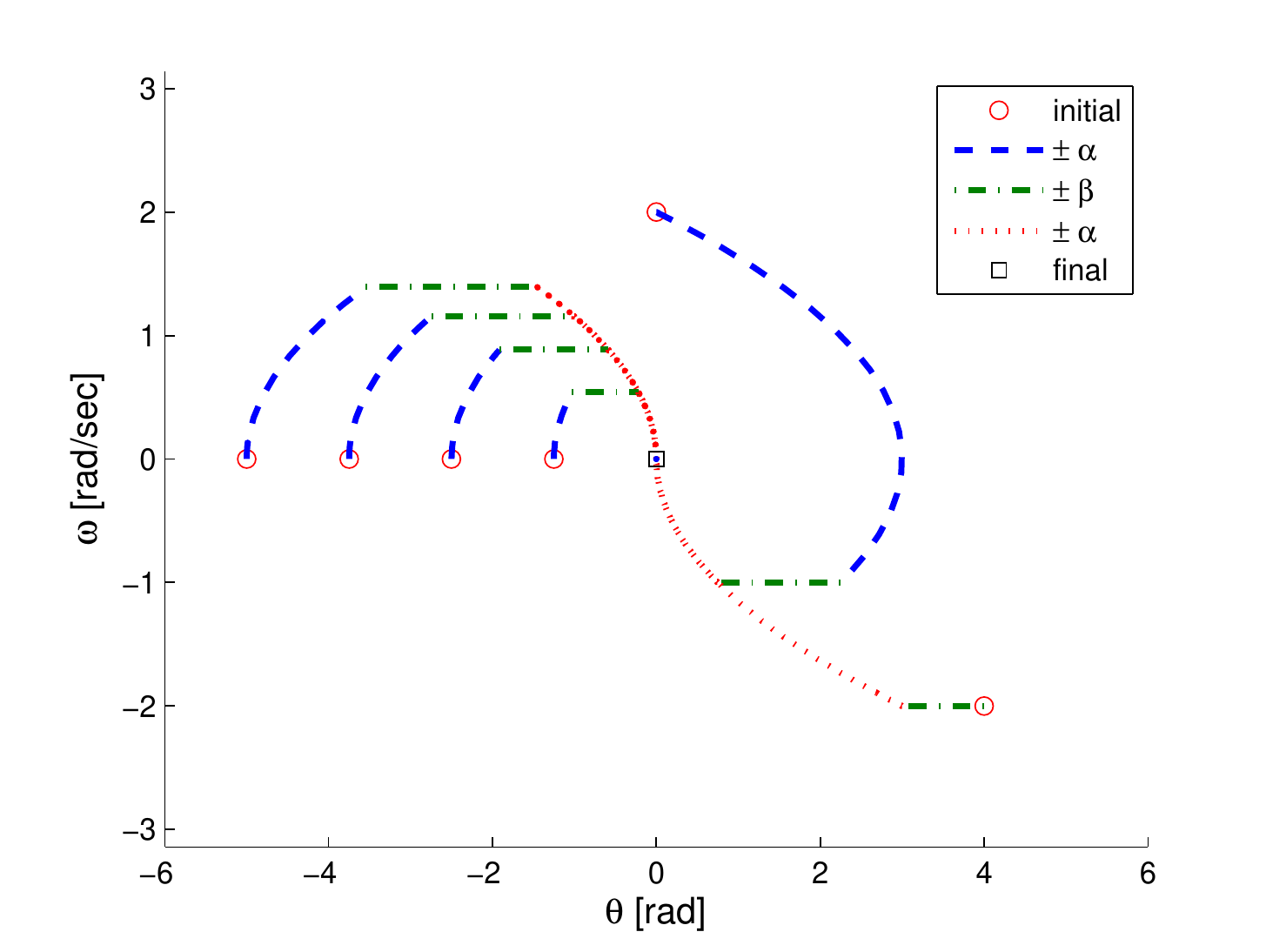}
\caption{Closed loop trajectories using \eqref{eq:statebasedcontrol1} for initial conditions in Figure \ref{fig:phase} (red circles) plotted in the $\theta-\omega$ plane. All trajectories reach the origin. These trajectories are identical to the open-loop time-optimal trajectories in Figure \ref{fig:phase}.}
\label{fig:simfb1}
\end{figure}

\begin{figure}[tb]
\centering
\includegraphics[width=0.4\textwidth]{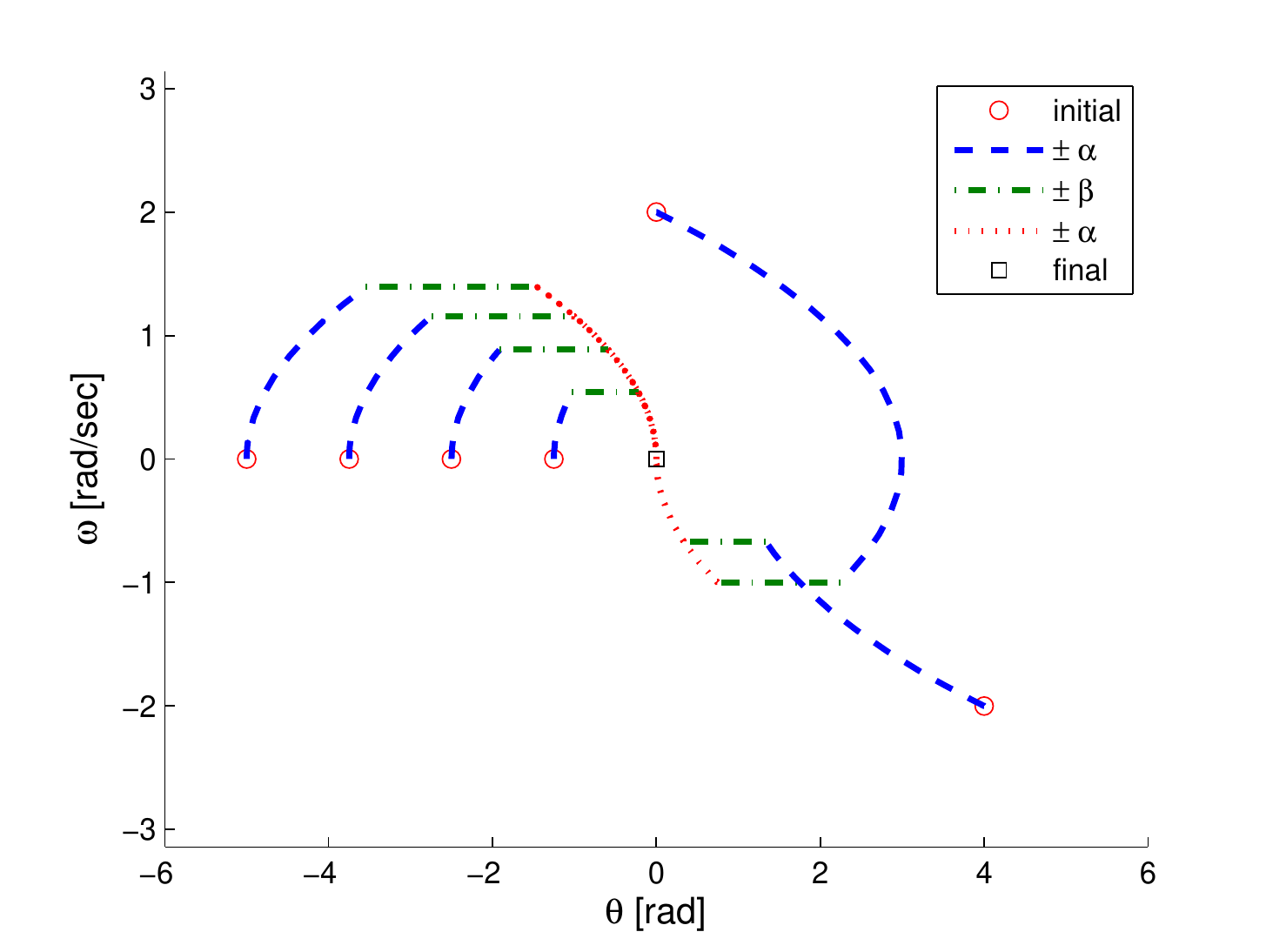}
\caption{Closed loop trajectories using \eqref{eq:statebasedcontrol2} for initial conditions in Figure \ref{fig:phase} (red circles) plotted in the $\theta-\omega$ plane. All trajectories reach the origin.}
\label{fig:simfb2}
\end{figure}

For the same initial conditions, instead of implementing open loop controls expressed as functions of time, we can use the state-based feedback controls \eqref{eq:statebasedcontrol1} and \eqref{eq:statebasedcontrol1}. The results are plotted in Figures \ref{fig:simfb1} and \ref{fig:simfb2} respectively. We can see that the closed loop trajectories plotted in Figure \ref{fig:simfb1} are identical to the time-optimal trajectories plotted in Figure \ref{fig:phase}. The difference between the two feedback control laws can be seen in the resulting closed-loop trajectory for the initial condition $(1 m/s,4 rad,-2 rad/s)$. The first trajectory leaves the region $\Omega_4$ by reaching the surface $H1(v,\theta,\omega) = 0$ (see Figure \ref{fig:simfb1}) while the second trajectory instead reaches the surface $H2(v,\theta,\omega) = 0$ upon leaving $\Omega_4$ (see Figure \ref{fig:simfb2}). 


\section{Velocity Control through Wheel speeds}
Suppose that the robot is such that the 
\begin{enumerate}
\item The control interface accepts commanded wheel speeds
\item There is a maximum allowed commanded wheel speed
\item The commanded wheel speeds are obtained practically instantly
\end{enumerate} 

As mentioned in Section \ref{sec:intro}, These assumptions are satisfied by platforms such as the iRobot Create.  Since the commanded wheel speeds can be achieved instantaneously, the state is simply $\theta$ whose dynamics are simply
\begin{dmath}
\dot{\theta} = \omega
\label{eq:kinematicstaticvd}
\end{dmath}
\noindent and the forward speed $v$ can change instantaneously. 

The kinematics of the system are given by \eqref{eq:ddrkinematics}. The right and left wheel speeds ( $\pdr$, $\pdl$ respectively) are bounded. Thus, $| \dot{\phi}_R | \leq \dot{\phi}_{max}$ and  $| \dot{\phi}_L | \leq \dot{\phi}_{max}$ for some $\dot{\phi}_{max} > 0$. The maximum forward speed $v_{max}$ and the maximum angular velocity $\omega_{max}$ are given by
\begin{dmath}
{v_{max}= r \dot{\phi}_{max},\ \omega_{max}= \frac{  r}{b} \dot{\phi}_{max}}
\end{dmath}

The robot may have a current heading $\theta$ and forward speed $v$. Our goal is to control the robot such that $\theta \rightarrow \theta_d$ and the forward speed $v$ equals the desired forward speed $v_d$.

We first look at the case when the desired velocity has a constant heading.
\subsection{Constant desired heading}
A straightforward application of the \PMP\ shows that in order to change to a linear speed $\| \mathbf{v}_d \|$ with heading $\theta_d$ in least amount of time, we can simply rotate in the required direction at maximum angular velocity until the heading $\theta$ matches the desired value, and then change wheel speeds to achieve the desired forward speed $v_d$. 

To see this, notice that the dynamics \eqref{eq:kinematicstaticvd} is of the form $f(q,u) = 0$, and thus the adjoint equation is given by $\dot{\psi} =0$,where $\psi \in \R $. Thus, the adjoint state is always $\psi(t) = \psi(0)$. The Hamiltonian is simply
\begin{dmath}
H = -\mu + \psi \omega
\end{dmath}
\noindent and since $| \omega | \leq \omega_{max}$, we can see that $H$ is maximized by selecting
\begin{dmath}
\omega = \omega_{max} \mysign{\psi(t)}\\
= \omega_{max} \mysign{\psi(0)}
\end{dmath}
Thus, the extremal control consists of a constant angular velocity. Clearly, then, the time-optimal control to change the heading $\theta$ and forward speed $v$ is as mentioned above.

\subsection{Time varying desired heading and desired forward speed}
Suppose the desired velocity is time-varying. 
If the goal state $\mathbf{v}_d$ cannot be predicted, the \PMP\ cannot be applied to such a system. The problem of tracking a time-varying trajectory has been tackled in previous research work \cite{}. Some results even achieve exponential tracking. However, these works often do not account for saturation, and are not concerned with shortest-time paths. 

An example of a continuous-time velocity tracking controller for a differential drive wheeled mobile robot is given by:

\begin{dgroup}
\begin{dmath}
v = \| v_d \| \cos (\theta - \theta_d)
\end{dmath}
\begin{dmath}
\omega = - k_{\omega} (\theta - \theta_d)
\end{dmath}
\label{eq:continuouskinematic}
\end{dgroup}
\noindent and in the absence of saturations, $v \rightarrow v_d$ and $\theta \rightarrow \theta_d$. If these desired quantities are time varying, such that the rates of change are bounded, then the error in tracking is also bounded. The bound can be reduced by increasing $k_{\omega}$. We now investigate the effect of saturated wheel speeds.

\begin{figure}[tb]
\centering
\begin{tikzpicture}[scale = 0.75]
\draw[->] (-4,0)-- (4,0);
\draw[->] (0,-4)-- (0,4);
\draw (0,4) node[anchor=south east]{$\omega$} ;
\draw (4,0) node[anchor=north west]{$v$} ;
\draw[rotate=45,dashed,line width = 0.5mm,fill = blue!30,fill opacity=0.5] (-1.414,-1.414) rectangle (1.414,1.414);
\draw (0,2) node[anchor=south east]{$\omega_{max}$} ;
\draw (2,0) node[anchor=north west]{$v_{max}$} ;
\end{tikzpicture}
\caption{The shaded region represents the achievable forward and angular velocities of the differential drive wheeled mobile robot}
\label{fig:vwphase}
\end{figure}
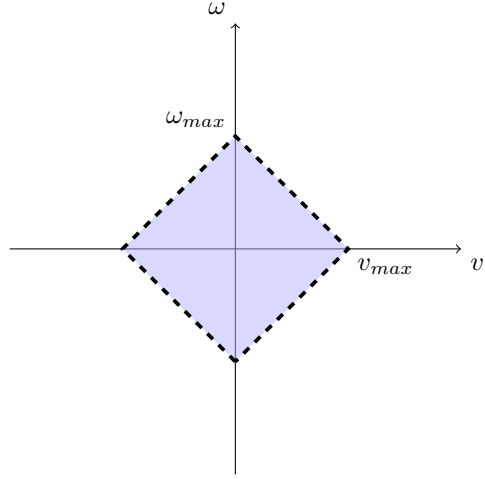

Due to the limits on the wheel speed, we can use \eqref{eq:ddrkinematics} to determine that the achievable forward and angular velocities are constrained to satisfy the relation
\begin{dmath}
\frac{| v |}{r} + \frac{b | \omega |}{r} \leq  \dot{\phi}_{max}
\label{eq:constrainedvw}
\end{dmath}
\noindent which is represented as the shaded region in Figure \ref{fig:vwphase}. 

Given the desired values of $v$ and $\omega$ from \eqref{eq:continuouskinematic}, the desired wheel speeds may be computed as 
\begin{dmath*}
\bmat{\dot{\phi}_{R,d}\\ \dot{\phi}_{L,d}}  = \frac{2 b}{ r^2} \bmat{-\frac{r b}{2} & -\frac{r}{2}\\ -\frac{r b}{2} & \frac{r}{2} } \bmat{v \\ \omega}
\end{dmath*}
If these wheel speeds are commanded, then the actual wheel speeds obtained are 
\begin{dmath*}
{\dot{\phi}_{R} = \mathrm{sat}(\dot{\phi}_{R,d} ,  \dot{\phi}_{max}) ,\dot{\phi}_{L} = \mathrm{sat}(\dot{\phi}_{L,d}, \dot{\phi}_{max} )} 
\end{dmath*}
\noindent where 
\begin{dmath}
\mathrm{sat}(x,\alpha) = \begin{cases} x & \mbox{ if } |x| < \alpha \\ \mysign{x} \alpha & \mbox{ otherwise }\end{cases} 
\end{dmath}

The actual forward velocity $v_{out}$ and $\omega_{out}$ achieved by the robot can be computed from these saturated wheel velocities using \eqref{eq:ddrkinematics}. They must satisfy \eqref{eq:constrainedvw}.

Thus, if $\| v_d\| > v_{max}$ then the heading angle does not converge. Let $\| v_d\| = v_{max} + \epsilon$, where $\epsilon > 0$. If $| \omega | < \frac{2 \epsilon}{b}$ then both desired wheel speeds are still greater than $\dot{\phi}_{max}$, implying that the actual forward and angular velocities due to the saturated wheel speeds will be $\pm v_{max}$ and $0$ respectively. Thus, when the error is non-zero yet small, the angular velocity remains zero, and the heading will not converge.

One can immediately see that this situation can be remedied by saturating the magnitude of the desired velocity. That is:
\begin{dgroup}
\begin{dmath}
v = \mathrm{sat}(\| v_d \| , v_{max}) \cos (\theta - \theta_d)
\end{dmath}
\begin{dmath}
\omega = - k (\theta - \theta_d)
\end{dmath}
\end{dgroup}
\noindent which allows the robot heading to converge, since $\epsilon = 0$ and hence the angular velocity after saturation is never $0$, unless the desired angular velocity is zero. 

Given that our goal is to re-orient the robot to match a desired heading and speed, the above method of computing $v$ and $\omega$ can be improved upon. The idea is to recognize that the achievable forward and angular velocities satisfy \eqref{eq:constrainedvw}. Then, given $v$ and $\omega$, we can compute $\bar{v}$ and $\bar{\omega}$ as 
\begin{dmath}
\bar{\omega} = \begin{cases} \mathrm{sign}(\omega_d) \omega_{max} & \mbox{ if } | \omega_d | \geq \omega_{max} \\ \omega_d &\mbox{ if } | \omega_d | < \omega_{max} \end{cases}
\end{dmath}
\noindent and 
\begin{dmath}
\bar{v} = \begin{cases} \mysign{v} \mathrm{max} \left( 0,  v_{max} -  b | \omega_d | \right)  & \mbox{ if } \frac{| v |}{r} + \frac{b | \omega |}{r}  >   \dot{\phi}_{max} \\
 v & \mbox{otherwise} \end{cases}
\end{dmath}
The effect is to always prioritize rotational motion when given $v$ and $\omega_d$ outside the shaded region in \ref{fig:vwphase}. 

A further improvement may be obtained through a heuristic solution that can be viewed as a hybrid control. Suppose that the desired heading is $\theta_d(t)$ and $|\dot{\theta}_d (t)|< \omega_{max}$. A bang-bang control can be used until $\theta(t) = \theta_d(t)$. After this time, the modified continuous tracking controller can be used. The benefit of this heuristic is that the lack of forward motion during the bang-bang phase minimizes the drift in position.

\subsection{Simulations}
\begin{figure}[tb]
\centering
\includegraphics[width=0.35\textwidth]{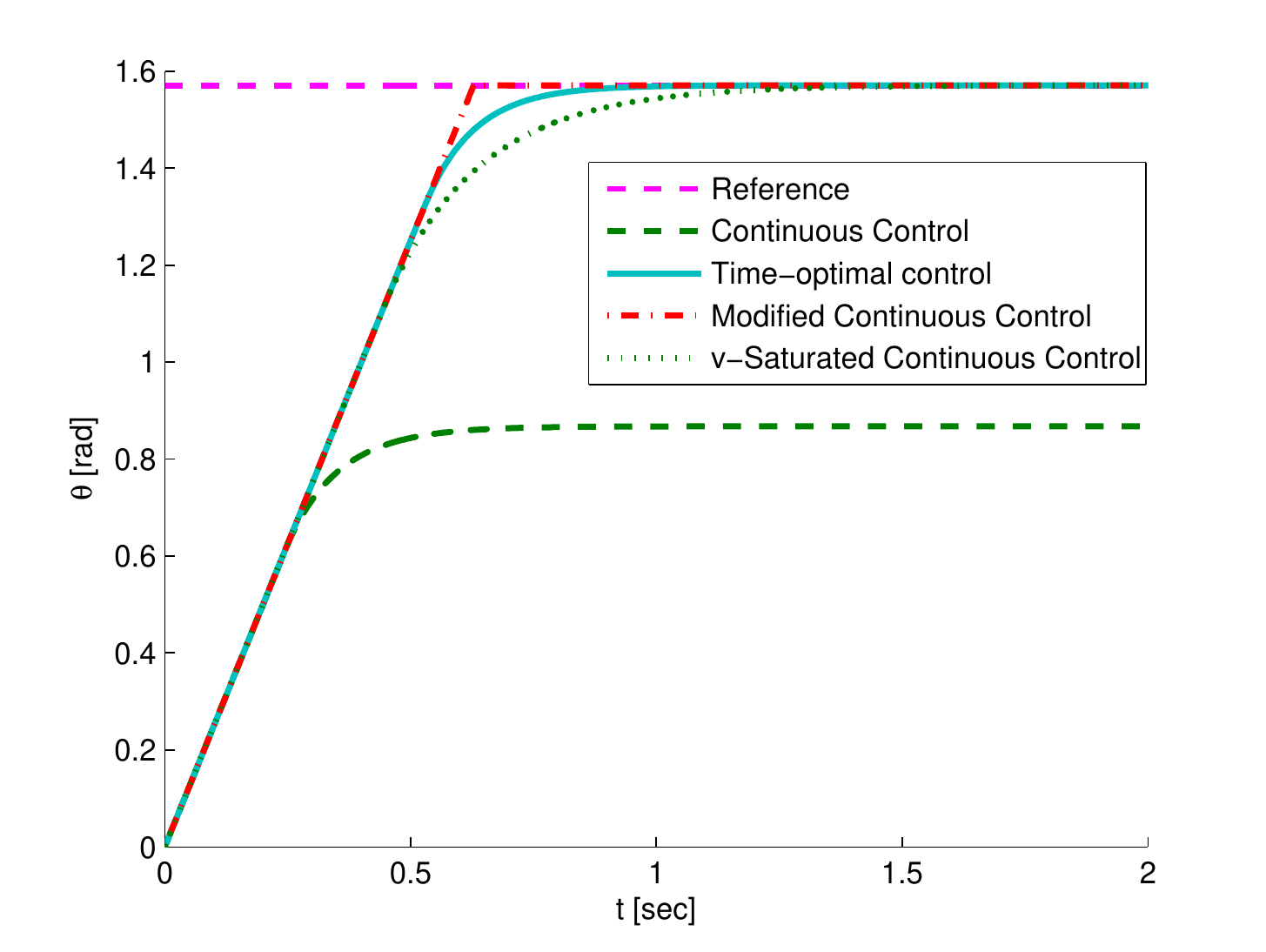}
\caption{Plot of $\theta$ versus time for different control strategies with a static desired velocity $\mathbf{v}_d$.}
\label{fig:vwphase1}
\end{figure}

Consider a robot with heading angle $0$. Let $r = 1 m$, $b = 5m$, and $\dot{\phi}_{max} = 0.5 rad/sec$. The robot is commanded to move with a velocity of $5 m/s$ in the positive $y$-axis direction. This implies that $\| \mathbf{v}_d \| > v_{max}$, and $\theta_d = \frac{\pi}{2}$. The results for the time-optimal control, the continuous control (without saturation of $\| \mathbf{v}_d \|$), the continuous control (with saturation of $\| \mathbf{v}_d \|$) and our the modified continuous control is given in Figure \ref{fig:vwphase1}. 

\begin{figure}[tb]
\centering
\includegraphics[width=0.35\textwidth]{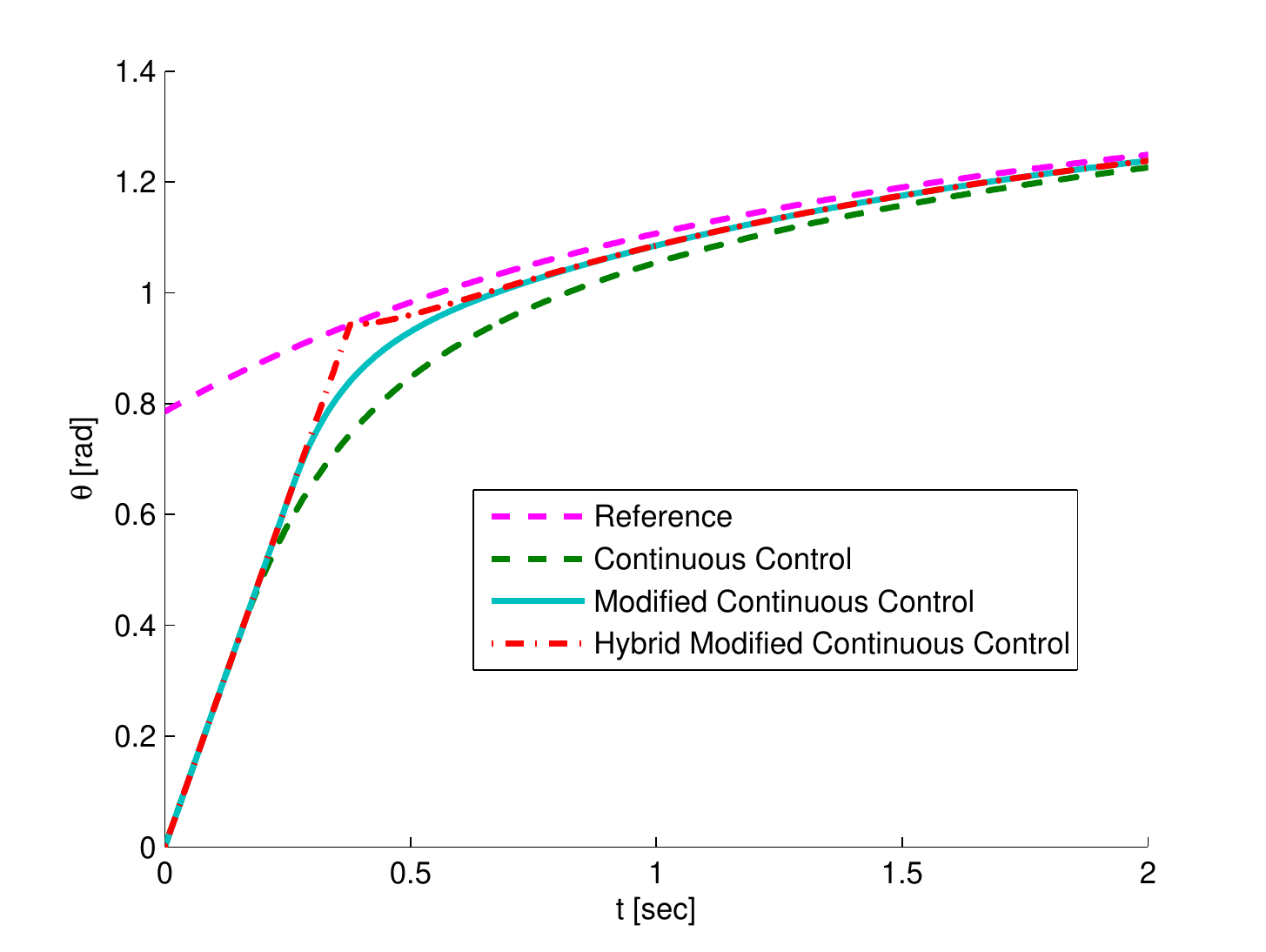}
\caption{Plot of $\theta$ versus time for different control strategies with a time-varying desired velocity $\mathbf{v}_d$.}
\label{fig:vwphase2}
\end{figure}

Next, we command a time-varying velocity $\mathbf{v}_d(t) = [1\ \ (1+t)]^T$. The results for the the continuous control, the modified continuous control (with saturation on $\mathbf{v}_d$) and the hybrid control strategy is given in Figure \ref{fig:vwphase2}. The modified and hybrid continuous controls definitely have lesser error as when compared to the continuous control case.
\section{Conclusion}
We have derived time-optimal controls that enable a torque-controlled differential drive wheeled mobile robot to reach a desired constant velocity in the plane in minimum time, for any initial velocity. These controls can be implemented as functions of time (planning problem) or as a feedback control based on a state-based switching rules.

\section{Acknowledgements}
The authors wish to thank Prof. Oleg Makarenkov at the University of Texas at Dallas for his helpful discussions. The first author became familiar with the details of the Pontryagin Maximum Principle thanks to Prof. Makarenkov's course on Switched Systems in the Spring of 2015.
\bibliographystyle{IEEEtran}
\bibliography{optimalcontrol,diffdrive}
\appendix
\label{sec:appendix}
\section{Extremals for goal states with non-zero angular velocity}
\label{app:wdnonzero}
We have presented a suitable feedback law that drives the differential drive robot from any initial forward speed $v$, angular velocity $\omega$ and heading $\theta$ to a desired forward speed $v_d$ and heading $\theta_d$, where the desired angular velocity is zero. The feedback law was derived after analyzing the set of extremals from any initial conditions that reached the desired goal.


In what follows, given initial and goal states, we can determine whether a $C1_{ns}$ or $C2$ control exists that results in a transition from the initial state to the goal state.

Consider an initial condition $(v,\theta,\omega)$. We can apply a control consisting of a sequence of three control phases of duration $t_1$, $t_2$ and $t_3$ respectively. Note that any of $t_1$, $t_2$ and $t_3$ may be zero. The first control phase consists of $\myNP$ or $\myPN$, with control  $\myNP$ or $\myPN$ in the third phase. The second phase consists of  $\myPP$ or $\myNN$ control. 

We can solve the time evolution of the state due to such a control quite easily as follows.

\begin{dgroup}
\begin{dmath}
v (\bar{t}_3) = v + s_2 \beta t_2
\end{dmath}
\begin{dmath}
\theta(\bar{t}_3)  = \theta +\omega t_1 + \frac{1}{2} s_1 \alpha t_1^2+ \omega(\bar{t}_1) t_2 + \omega(\bar{t}_2) t_3 + \frac{1}{2} s_3 \alpha t_3^2
\end{dmath}
\begin{dmath}
\omega(\bar{t}_3)  = \omega +  s_1 \alpha t_1  + s_3 \alpha t_3
\end{dmath}
\end{dgroup}
\noindent where $s_1,s_2, s_3 \in \{1,-1\}$ and $\bar{t}_2 = t_1 + t_2$, $\bar{t}_3 = t_1+t_2+t_3$. If the first (third) control phase is $\myNP$, then $s_1 = 1$ ($s_3 = 1$) , otherwise $s_1 = -1$ ($s_3 = -1$). If the second control phase is $\myPP$, then $s_2 = 1$, otherwise $s_2=-1$. 

We wish to solve for $t_1$, $t_2$, $t_3$ such that 

\begin{dgroup*}
\begin{dmath*}
{t_1 \geq 0, \hspace{5mm} t_2 \geq 0, \hspace{5mm} t_3 \geq 0}
\end{dmath*}
\begin{dmath*}
{v(\bar{t}_3) = 0, \theta(\bar{t}_3) = 0,  \omega(\bar{t}_3) = \omega_d} 
\end{dmath*}
\end{dgroup*}

We can immediately solve for $t_2$ and $s_2$:
\begin{dmath}
t_2 = \frac{-v}{s_2 \beta}\\
	= \frac{|v|}{\beta}
	\label{eq:t2}
\end{dmath}
\noindent where $s_2 = \mysign{-v}$. We can then express $t_3$ in terms of $t_1$ as 
\begin{dmath}
t_3 = -\frac{s_1}{s_3} t_1 + \frac{\omega_d - \omega}{s_3 \alpha}
	\label{eq:t3}
\end{dmath}
Which we can substitute into the expression for $\theta(\bar{t}_3)$ as 
\begin{dmath}
\theta(\bar{t}_3) = \theta(\bar{t}_2)+ \omega(\bar{t}_2) t_3 + \frac{1}{2} s_3 \alpha  t_3^2\\
= \theta(t_1) + \omega(t_1) t_2+ \omega(\bar{t}_2) t_3 + \frac{1}{2} s_3 \alpha  t_3^2\\
= \theta + \omega t_1 + \frac{1}{2} s_1 \alpha  t_1^2 + \omega(t_1) t_2+ \omega(\bar{t}_2) t_3 + \frac{1}{2} s_3 \alpha  t_3^2\\
= \theta + \omega t_1 + \frac{1}{2} s_1 \alpha  t_1^2 + \left( \omega + s_1 \alpha t_1 \right) \frac{|v|}{\beta}+ \left(\omega + s_1 \alpha t_1 \right) \left( -\frac{s_1}{s_3} t_1 + \frac{\omega_d - \omega}{s_3 \alpha} \right) - \frac{1}{2} s_1 \alpha \left( -\frac{s_1}{s_3} t_1 + \frac{\omega_d - \omega}{s_3 \alpha} \right)^2\\
= s_1 \alpha t_1^2 + \bigpar{2 \omega + \frac{s_1 \alpha |v|}{\beta}} t_1 + \bigpar{ \theta + \frac{\omega |v|}{\beta}} + \frac{\omega^2 - \omega_d^2}{2 s_1 \alpha}
\end{dmath}

Once we set the $\theta(\bar{t}_3)$ to be equal to the desired value of zero, we obtain
\begin{dmath}
\frac{s_1-s_3}{2} \alpha t_1^2 + \bigpar{ \omega \frac{(s_3-s_1)}{s_3} + \frac{s_1 \alpha |v|}{\beta}} t_1 + \bigpar{ \theta + \frac{\omega |v|}{\beta}} + \frac{\omega_d^2 - \omega^2}{2 s_3 \alpha} = 0
\label{eq:t1}
\end{dmath}

If $s_1 = -s_3$ then the equation is quadratic with solutions 

\begin{dgroup}
\begin{dmath}
t_1  = \frac{1}{2 s_1 \alpha} \bigpar{ - \bigpar{2 \omega + \frac{s_1 \alpha | v | }{\beta}} \pm \sqrt{\Delta} }
\end{dmath}
\begin{dmath}
t_3  = \frac{1}{2 s_1 \alpha} \bigpar{ - \bigpar{2 \omega_d + \frac{s_1 \alpha | v | }{\beta}} \pm \sqrt{\Delta} }
\end{dmath}
\begin{dmath}
\Delta  = 2 \omega^2 + 2 \omega_d^2 + \frac{\alpha^2 | v |^2 }{\beta^2} - 4 s_1 \alpha \theta
\end{dmath}
\label{eq:alphabetaalphaquadsol}
\end{dgroup}
Thus, there are four solutions for $t_1$ and $t_3$ (two for $s_1=1$ and two for $s_1=-1$). Not all solutions may be such that both $t_1$ and $t_3$ are non-negative. The total time for any valid solution can be computed to be
\begin{dmath}
\bar{t_3} = t_1 + t_2 + t_3\\
= \pm \sqrt{\Delta} - \frac{\omega_d + \omega}{s_1 \alpha}
\end{dmath}

Instead, if $s_1 = s_3$, then \eqref{eq:t1} reduces to a linear equation. The solution is
\begin{dgroup}
\begin{dmath}
t_1  = \frac{\beta (\omega^2-\omega_d^2)}{2 \alpha^2 |v|} - \frac{\theta \beta}{s_1 \alpha |v| } - \frac{\omega}{s_1 \alpha}
\end{dmath}
\begin{dmath}
t_3  = \frac{\beta (\omega_d^2-\omega^2)}{2 \alpha^2 |v|} + \frac{\theta \beta}{s_1 \alpha |v| } + \frac{\omega_d}{s_1 \alpha}
\end{dmath}
\begin{dmath}
s_1 = \mysign{\omega_d - \omega}
\end{dmath}
\label{eq:alphabetaalphalinsol}
\end{dgroup}
\noindent for which there is only one possible solution. Thus, we have five possible solutions for $(t_1, t_2, t_3)$.
\begin{figure}[tb]
\centering
\includegraphics[width=0.4\textwidth]{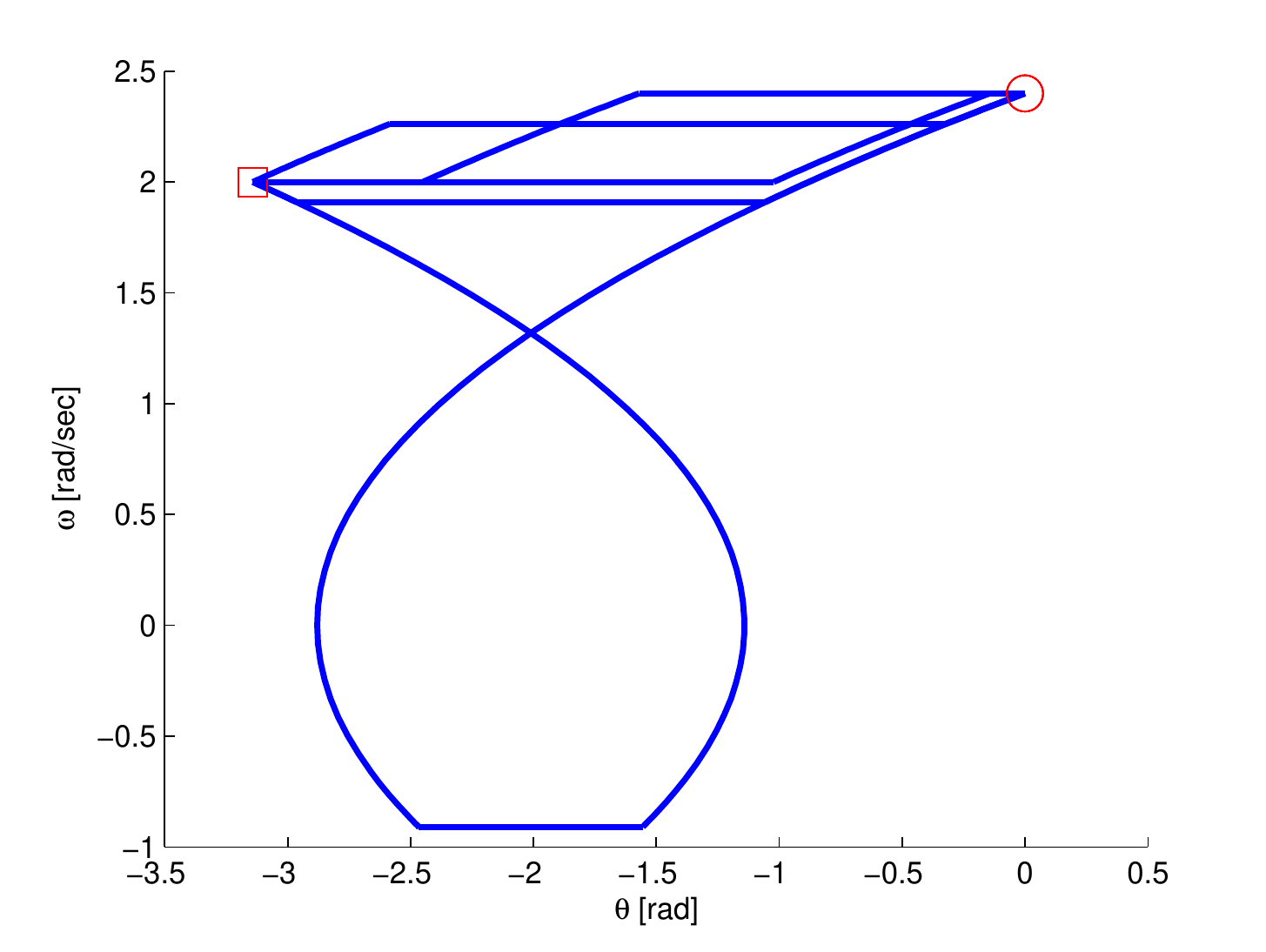}
\caption{Five extremals in the  $\theta - \omega$ plane for initial condition  $q_0 = (3,-\pi,2)$ (red square) and goal state $q_d = (0,0,2.4)$ (red circle). }
\label{fig:wdnonzero}
\end{figure}
We follow a similar procedure when  the first and third control phases consists of  $\myPP$ or $\myNN$ control, and the second phase consists of  $\myNP$ or $\myPN$ control.  The integrated equations are:
\begin{dgroup}
\begin{dmath}
v (\bar{t}_3) = v + s_4 \beta t_1 + s_6 \beta t_3
\end{dmath}
\begin{dmath}
\theta(\bar{t}_3) = \theta + \omega t_1 + \omega(t_1) t_2 + \frac{1}{2} s_5 \alpha t_2^2+ \omega(\bar{t}_2) t_3
\end{dmath}
\begin{dmath}
\omega(\bar{t}_3) =\omega + s_5 \alpha t_2
\end{dmath}
\end{dgroup}

\noindent where $s_4, s_5, s_6 \in \{1,-1\}$ and $\bar{t}_2 = t_1 + t_2$, $\bar{t}_3 = t_1+t_2+t_3$. If the second control phase is $+\alpha$, then $s_5 = 1$, otherwise $s_5 = -1$. If the first control phase is $+\beta$, then $s_4 = 1$, otherwise $s_4=-1$. Similarly, if the third control phase is $+\beta$, then $s_6 = 1$, otherwise $s_6=-1$. 


We can compute
\begin{dmath}
t_2 = \frac{\omega_d - \omega}{s_5 \alpha}
\end{dmath}
\noindent which implies that $s_5 = \mysign{\omega_d - \omega}$.

Setting $v (\bar{t}_3) =0$ and $\omega(\bar{t}_3) = \omega_d$, we can obtain

\begin{dgroup}
\begin{dmath}
t_1 = \bigpar{\omega - \frac{s_4}{s_6} \omega_d}^{-1} \bigpar{\frac{\omega_d v}{s_6 \beta} + \frac{\omega^2 - \omega_d^2}{2 s_5 \alpha} - \theta}
\end{dmath}
\begin{dmath}
t_3 = \bigpar{\omega_d - \frac{s_6}{s_4} \omega}^{-1} \bigpar{\frac{\omega v}{s_4 \beta} + \frac{\omega^2 - \omega_d^2}{2 s_5 \alpha} - \theta}
\end{dmath}
\label{eq:finaltimesc1}
\end{dgroup}

The total time taken for such a solution is 
\begin{dmath}
\bar{t_3} = \begin{cases} -\frac{v}{s_4 \beta} + \frac{|\omega_d - \omega|}{\alpha} & \mbox{ if } s_4 s_6 = 1\\
\frac{\omega-\omega_d}{\omega+\omega_d} \frac{v}{s_4 \beta} - \frac{2 \theta}{\omega+\omega_d} & \mbox{ if } s_4 s_6 = -1
\end{cases}
\end{dmath}

Thus, given $(v,\theta, \omega)$, we can compute \eqref{eq:alphabetaalphaquadsol} when $s_1 \in \{1,-1 \}$, \eqref{eq:alphabetaalphalinsol} which has only one possible solution, and \eqref{eq:finaltimesc1} for four possible values of the pair ($s_4$,$s_6$). This results in nine values of the triplet $(t_1,t_2,t_3)$. The control corresponding to the least value of  $t_1+t_2+t_3$, where all three durations are non-negative, is selected as the time-optimal control.

Consider the intial condition $q_0 = (3 m/s,-\pi rad,2 rad/sec)$ and goal state $q_d = (0 m/s,0 rad,2.4 rad/sec)$ when $\alpha = 2/3$, $\beta=2/3$. There are five extremals that achieve the transition from $q_0$ to $q_d$, and only one is time-optimal. These extremals are plotted in the $\theta - \omega$ plane in Figure \ref{fig:wdnonzero}. The initial condition $q_0 = (-1 m/s,-\pi rad,4 rad/sec)$ with goal state $q_d = (0 m/s,0 rad,4.4 rad/sec)$ also has five extremal solutions, however two of them are time-optimal. A future goal is to propose a feedback control law for the case when the goal angular velocity is non-zero, as was done for the case when it is zero.

\section{Theorem 4, $\S$10, \cite{Filippov88}}
\label{app:filippv}
\noindent Condition 1:\hfill\\

Let a domain $G \subset \R^3$be separated by smooth hypersurfaces $s_i^k$ into domains $S_j^n$, $j = 1,\dots, r$. The superscript denotes the dimension of the surface, the subscript denotes the index of the surface or domain. The boundary of each hypersurface does not belong to the surface, and consists of a finite number of smooth hypersurfaces of smaller dimensions or points. 

The vector valued function $f(t,q)$ is continuous in $t$,$q$ for $a < t < b$ in each of the domains $S_j^n$ upto the boundary, that is, $f(t,q) = f_j^n(t,q)$ for $q \in S_j^n$, and the function $f_j^n$ is continuous in $\bar{S}_j^n$. On sum or all of the hypersurfaces $\bar{S}_j^k$, $0 \leq k \leq n-1$, or on some of their closed subsets continuous vector valued functions $f_i^k(t,q)$ are given; the vector $f_i^k(t,q)$ lies in the $k$-dimensional plane tangent to $S_i^k$ at the point $q$.
\end{document}